\documentclass[amssymb, nofootinbib, superscriptaddress]{revtex4}

\usepackage{amsmath}
\usepackage{amsfonts}
\usepackage{graphicx}
\usepackage{psfrag}
\usepackage{array}
\usepackage{bbm}
\usepackage{subfigure}
\usepackage{float}
\usepackage[dvips]{color}

\newcommand{\f}{\frac}
\newcommand{\tr}{\mathrm{tr}}
\newcommand{\rar}{\rightarrow}

\newcommand{\su}{\mathfrak{su}}
\newcommand{\SU}{\mathrm{SU}}
\newcommand{\SO}{\mathrm{SO}}

\newcommand{\C}{\mathbb{C}}
\newcommand{\N}{\mathbb{N}}
\newcommand{\Z}{\mathbb{Z}}

\newcommand{\cA}{{\cal A}}
\newcommand{\cB}{{\cal B}}

\newcommand{\cD}{{\cal D}}
\newcommand{\cF}{{\cal F}}
\newcommand{\cG}{{\cal G}}

\newcommand{\cM}{{\cal M}}

\newcommand{\cO}{{\cal O}}
\newcommand{\cS}{{\cal S}}
\newcommand{\cT}{{\cal T}}
\newcommand{\cZ}{{\cal Z}}

\newcommand{\id}{\mathbb{I}}

\newcommand{\Str}{\textrm{Str}}
\newcommand{\Sdim}{\textrm{Sdim}}
\newcommand{\Schi}{\textrm{S}\chi}

\newcommand{\be}{\begin{equation}}
\newcommand{\ee}{\end{equation}}
\newcommand{\bes}{\begin{eqnarray}}
\newcommand{\ees}{\end{eqnarray}}

\newcommand{\ua}{\uparrow}
\newcommand{\da}{\downarrow}

\newcommand{\w}{\wedge}
\newcommand{\osp}{\mathfrak{osp}}
\newcommand{\OSP}{\textrm{OSP}}
\newcommand{\uosp}{\mathfrak{uosp}}
\newcommand{\UOSP}{\textrm{UOSP}}

\newcommand{\Ref}[1]{(\ref{#1})}

\begin{document}
%
\title{\Large\bf The particle interpretation of N=1 supersymmetric spin foams}

\author{{\bf Valentina Baccetti}}\email{baccetti@neve.fis.uniroma3.it}
\affiliation{Dipartimento di Fisica "E. Amaldi", Universit\`a degli Studi Roma Tre, Via della Vasca Navale 84, 00146 Roma, Italy }

\author{{\bf Etera R. Livine}}\email{etera.livine@ens-lyon.fr}
\affiliation{Laboratoire de Physique, ENS Lyon, CNRS UMR 5672, 46 All\'ee d'Italie, 69007 Lyon, France}

\author{{\bf James P. Ryan}}\email{james.ryan@aei.mpg.de}
\affiliation{MPI f\"ur Gravitationsphysik, Albert Einstein Institute,  Am M\"uhlenberg 1, D-14476 Potsdam, Germany}

\date{\small \today}


\begin{abstract}

We show that  $N=1$-supersymmetric $BF$ theory in 3d leads to a supersymmetric spin foam amplitude via a lattice discretisation.   Furthermore, by analysing the supersymmetric quantum amplitudes, we show that they can be re-interpreted as 3d gravity coupled to embedded fermionic Feynman diagrams.

\end{abstract}

\maketitle

\section{Introduction}
\label{intro}
Whilst supergravity theories go a certain way to tame the infinities of their non-supersymmetric cousins, they also generalise these theories by coupling fermionic degrees of freedom to the original gravity theory.  One can easily see this at the continuum classical level. We want to investigate this issue in the discrete quantum case. Thus, in this paper, we shall analyze a rather simplified model: Riemannian supergravity in three dimensions.  As is well known, the conventional metric-dependent action may be recast as particular gauge theory action, which is dependent on the triad and the spin connection: $BF$ theory.  Extensive work on the discretisation and quantisation of this theory with $\SU(2)$ chosen as the gauge group arises in the literature (see \cite{naish, PR1} and references therein).  One can see that the theory maintains its topological nature once quantised, and the discrete quantum model is the Ponzano-Regge model.   Several approaches to coupling matter within spin foams were embarked upon \cite{PR1, livineoeckl, winston1, winston2, YM, Livine:2007dx}.  The most tractable and indeed most successful of these procedures embedded the Feynman diagrams of the field theory into the spin foam.  Remarkably, summing over the gravitational degrees of freedom, the effective matter amplitude was seen to arise as the Feynman diagram of a non-commutative field theory \cite{noncom}.  To add to this position, it was shown that an explicit 2nd quantised theory of this gravity matter theory could be provided by group field theory, while later the non-commutative field theory was seen to arise as a phase around a classical solution of a related group field theory \cite{gftmat}.  Of course, one may approach the subject with the view that one should discretise the field directly on the spin foam, since in the continuum theory, we expect that the field has a non-trivial energy-momentum tensor, and should affect the state sum globally.  This method has yielded to a succinct initial quantisation for Yang-Mills and fermionic theories \cite{winston1, winston2, YM}, but due to the non-topological nature of the resulting amplitudes, further calculations proved unwieldy.  Now, it was not our intention that this work would or should settle this debate, but we find that this theory is more in line with the arguments of the former way.

The path we follow in our analysis is to start from continuum $BF$ theory with gauge group $\UOSP(1|2)$, discretise and quantise. Once this lattice gauge theory quantisation has been completed, we Fourier transform to $\uosp(1|2)$ representation space.  Owing to its algebraic structure, there is an $\su(2)$ structure embedded within $\uosp(1|2)$ \cite{Scheunert:1976wj, bertol, Daumens:1992kn}.  We may rewrite the amplitudes to make this dependence explicit.  The aim of the game is then to give an accurate interpretation of these amplitudes in terms of matter coupled to gravity.  To do so,  we Fourier transform again, but this time to functions on the group $\SU(2)$.  In this form, we can identify within the state sum Feynman diagrams of a massless spin-$\frac{1}{2}$ fermionic field.   Therefore, we arrive at a lattice discretisation of gravity coupled to a fermionic field where the supersymmetric nature of the theory is hidden.

\section{The Super Ponzano-Regge model}
\label{derivation}

In 3d, we can rewrite the gravity action (with zero cosmological constant) as an $\SU(2)$ gauge theory:
\be
\cS[E,W]=\int_{\cM} \tr(E\w F[W]),\label{class1}
\ee
where $E$ is the triad, $W$ is the connection, while $F[W] =  dW + W\w W$ is the curvature.  Both $E$ and $W$ are 1-forms valued in $\su(2)$, and $\tr$ is the trace over the Lie algebra: $\tr(L_iL_j)  = -\frac{1}{2}\delta_{ij}$.\footnotemark
\footnotetext{The $\su(2)$ algebra has generators satisfying: $[L_i,L_j] =  -\epsilon_{ijk}L_k$, and in the fundamental representation has the form:
\be
L_1 = \frac{i}{2}
\left(\begin{array}{cc}
0 & 1\\
1 & 0
\end{array}
\right),
\quad\quad\quad\quad
L_2 = \frac{i}{2}
\left(\begin{array}{cc}
0 & -i\\
i & 0
\end{array}
\right),
\quad\quad\quad\quad
L_3 = \frac{i}{2}
\left(\begin{array}{cc}
1 & 0\\
0 & -1
\end{array}
\right).\label{class2}
\ee}
In this paper, we update the analysis to $3d$ supergravity with zero cosmological constant:
\be
\cS[\cB,\cA]=\int_{\cM} \Str(\cB\w\cF[\cA]),\label{class3}
\ee
where $\cB$ is the supertriad, $\cA$ is the superconnection, while $\cF[\cA] = d\cA + \cA \w \cA$ is the supercurvature.  $\Str$ is its supertrace. The minimal supersymmetric extension of $3d$ gravity is to take the gauge algebra: $\uosp(1|2)$.  This type of theory was first conceived in the context of non-zero cosmological constant where it is equivalent to a super Chern-Simons theory devised by Ach\'ucarro and Townsend \cite{aapt}.

The supergravity fields written in terms of generators of the algebra are
\be
\cB = E^iJ_i + \phi^A Q_A, \qquad \cA = W^iJ_i + \psi^A Q_A,\label{class4}
\ee
where $E$ and $W$ are the triad and connection, while $\phi$ and $\psi$ represent the fermion
field.   $A\in\{\pm\}$ and $i\in\{1,2,3\}$.  The action may be rewritten in terms of these
variables as
\be
\cS_{N = 1}[E,W,\phi,\psi] = \int_{\cM} \Str\Big(E\w (F[W] + \psi\wedge\psi) + \phi \w d_W\psi\Big),\label{class5}
\ee
where $F(W) = dW + W\w W$ is the gravitational curvature, and we define the operator as $d_W = d +
W\w$.  This action describes a fermion field propagating on a manifold $\cM$ endowed with a
dynamical geometry.

Let us conclude our classical analysis by saying a few words on the equations of motion.  In supersymmetric form they are:
 \be
\cF[\cA] = 0 \quad\quad \textrm{and} \quad\quad d_\cA \cB = 0.\label{class6}
 \ee
Thus, they record that the classical solutions satisfy the condition that  the superconnection is super-flat and super-torsion-free.  Breaking this up into components we see that:
\bes
F[W]^i + \frac{i}{2}(\sigma^i)_{AB}\, \psi^A \wedge \psi^B & = & 0, \label{class7}\\
d\psi^A + \frac{i}{2} (\sigma_i)_B{}^A \,W^i\wedge \psi^B  & = & 0,\label{class8}\\
d_W e^i+ \frac{i}{2}(\sigma^i)_{AB}\, \psi^A \wedge \phi^B & = & 0,\label{class9}\\
d\phi^A + \frac{i}{2}(\sigma_i)_B{}^A\,  W^i \wedge \phi^B + \frac{i}{2}(\sigma_i)_B{}^A\,  e^i \wedge \psi^B & = & 0,\label{class10}
\ees
where $(\sigma_i)_A{}^B$ are the Pauli matrices.  Going in descending order, we see that the curvature of the $\su(2)$ connection is non-vanishing, it picks up a contribution from the matter sector \eqref{class7}.  Furthermore, the fermion field, acting as a source for the curvature, is covariantly constant with respect to the $\su(2)$ connection \eqref{class8}.   Equation \eqref{class9} states that $\su(2)$ connection is not torsion free, while \eqref{class9} and \eqref{class10} together, show that any change in the triad is compensated by a change in the fermionic fields.   This complementary viewpoint shall come into play later in Section \ref{susyone}.

\begin{figure}[H]
\centering
\psfrag{a}{$v^*$}
\psfrag{b}{$e^*$}
\psfrag{c}{$v$}
\psfrag{d}{$e$}
\psfrag{w}{$w^*$}
\psfrag{e}{$e^*_t$}
\psfrag{f}{$v^*_e$}
\psfrag{g}{$v^*_f$}
\psfrag{h}{$e^*_{e,f}$}
\begin{center}
\includegraphics[width = 7cm]{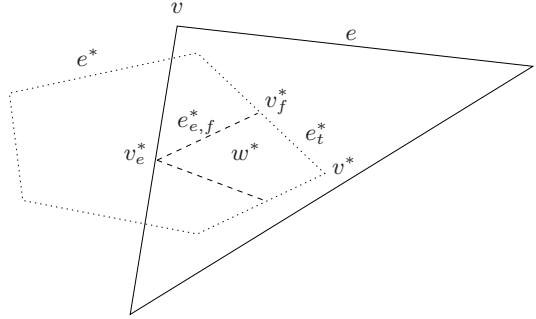}
\caption{\label{wedge}The various elements of the simplicial complex and its dual.}
\end{center}
\end{figure}

The aim of the game is to rigorously define the path integral:
\be
Z_{\cM} = \int\cD\cA\,\cD\cB \; e^{i\cS_{N=1}[\cB, \cA]}\quad \textrm{\lq\lq$\,=\,$\rq\rq}\,  \int \cD\cA \; \delta(\cF[\cA]). \label{quant1}
\ee
In order to embark upon our course of discretisation and quantisation, we must first introduce a number of structures:
\begin{itemize}
\item  We replace the manifold $\cM$ by a simplicial manifold $\Delta$ of the same topology. Since $BF$ theory is a topological field theory and does not have any local degrees of freedom, we expect that this substitution will preserve the information contained in the continuum theory.  In three space-time dimensions, we can triangulate any manifold. We label the 0-,1-,2- and 3-subsimplices as $v,e,f$ and $t$, respectively.

\item Another important constituent is the topological dual $\Delta^*$ to the simplicial complex.  We label sub-elements of this structure as $v^*,e^*,f^*$ and $t^*$, respectively.  Furthermore, the dual 2-skeleton $\Delta^*_2\subset\Delta$ is defined to be $\{v^*, e^*,f^*\}$.

\item  We subdivide the dual 2-skeleton $\Delta^*_2$.  The edges $e^*\in\Delta^*$ intersect the triangles $f\in\Delta$. We label these points of intersection by $v^*_f$.  Likewise, the edges $e\in\Delta$ intersect the faces $f^*\in\Delta^*$.  We label these points of intersection by $v^*_e$.  Changing viewpoints, the vertices $v^*_f$ split the edges $e^*$ into two parts, which we label $e^*_t$. Furthermore, we join $v^*_f$ to $v^*_e$ which an edge labelled by $e^*_{e,f}$.  This allows us to demarcate the wedges $w^*\subset f^*$ which are circumscribed by a combination of edges $e^*_t$ and $e^*_{e,f}$ - the details are drawn in Fig. \ref{wedge}.  To summarise, a wedge is that part of a face $f^*$ contained in the interior of a single tetrahedron.

\end{itemize}

The fields $\cB$ and $\cA$ are replaced by configurations that are distributional with support on subsimplices of $\Delta$ and its topological dual $\Delta^*$. We integrate these fields over the appropriate subsimplices. The definition of the integrated fields is:
\be
\begin{array}{lcll}
\cB           & \rar &   \displaystyle B_{w^*}         = \int_{e\sim w^*} \cB                   & \in \uosp(1|2),\vspace{0.3cm}\\

\cA      & \rar & \displaystyle   g_{e^*}     = {\cal P} e^{\int_{e^*}\cA}      &  \in \UOSP(1|2),\vspace{0.3cm}\\
\cF[\cA]   & \rar & \displaystyle    G_{w^*}         =  \prod_{e^*\subset \partial w^*} g_{e^*}^{\epsilon_{(e^*,f^*)}}  & \in \UOSP(1|2),
\end{array}\label{quant2}
\ee
where $\epsilon_{(e^*,f^*)} = \pm1$ depending the relative orientation of $e^*$ and $f^*$.
The super-flatness constraint $\cF[\cA]=0$ translates into the triviality of holonomies:  $G_{w^*}=\id$.  At the discrete level, we expect to replace the $\delta(\cF[\cA])$ constraint by $\delta(G_{w^*})$ constraints, with the discrete $B_{w^*}$ variables still playing the role of the Lagrange multipliers. Following \cite{PR1}, the action on the simplicial manifold then reads:\footnote{Remember that: \be
\begin{array}{rclcrcl}
 \Str(BG)& = & b_i\, p_i + b^\square\, p^\square + b\, p, & \text{where}&\vec{p} & =& -\frac{1}{2}\left( 1 - \frac{1}{8}\eta^\square \eta\right)
 \left(\begin{array}{c}
 i\, (u_{32} + u_{23} )\vspace{0.1cm}\\
 u_{32} - u_{23}\vspace{0.1cm}\\
 i\,(u_{22} - u_{33})
 \end{array} \right)  \vspace{0.1cm} \\
 &&&\text{and}& p & = & -\frac{1}{4} \left( \eta^\square  + \eta^\square\, u_{22} + \eta\, u_{23} \right),
 \end{array}\label{quant3}
 \ee
 and the definitions of $u_{ij}$ are given in Appendix \ref{OSP}.}
\be
\cS_\Delta[B_{w^*}, g_{e^*_t}, g_{e^*_{e,f}}] = \sum_{w^*\in\Delta^*} \Str(B_{w^*}G_{w^*}).\label{quant4}
\ee
An important factor in rigorously defining the path integral is the measure:
\be{
\begin{array}{lclcrcl}

\cD\cA &\rightarrow& \displaystyle\prod_{e_t^*} d g_{e^*_t}\prod_{e^*_{e,f}}dg_{e^*_{e,f}} &\phantom{xxxxxx} \text{where}\phantom{xxxxxx} & dg & = & \frac{1}{\pi^2}(1-\frac{1}{4}\eta^\square\eta) \sin^2 \theta \sin \psi\; d\theta\, d\psi\, d\phi\, d\eta^\square\, d\eta, \vspace{0.2cm} \\

\cD\cB &\rightarrow& \displaystyle\prod_{w^*} d B_{w^*} &\phantom{xxxxxx} \text{where}\phantom{xxxxxx} & dB &=& db_1\,db_2\,db_3\,db^{\square} \,db,

\end{array}}\label{quant5}
\ee
where $\eta^\square,\, \eta,\, b^\square,\, b$ are odd Grassmann variables while the rest are even  and parameterise the $\SU(2)$ sub-group (we refer to Appendix \ref{OSP} for more details).   Thus, the path integral for a discrete manifold takes the form:
\be
\cZ_{\Delta, \UOSP(1|2)} = \int \prod_{e^*_t}  dg_{e^*_t}\prod_{e^*_{e,f}} dg_{e^*_{e,f}}\prod_{w^*} dB_{w^*}\;   e^{i\,\cS_\Delta[B_{w^*}, g_{e^*_t}, g_{e^*_{e,f}}]}.\label{quant6}
\ee
Our next step is to integrate over the supertriad:
\be
\begin{split}
&\cZ_{\Delta, \UOSP(1|2)}
= \int \prod_{e^*_t} dg_{e^*_t} \prod_{e^*_{e,f}}dg_{e^*_{e,f}} \prod_{w^*}\;(p^\square_{w^*}\, p_{w^*})\,\delta^{3}(\vec{p}_{w^*}) \\
&\phantom{xxxxxxxxxxxxxxxxxxx}
=  \int \prod_{e^*_t} dg_{e^*_t} \prod_{e^*_{e,f}}dg_{e^*_{e,f}} \prod_{w^*} \; \left[\delta(p^\square_{w^*})\, \delta(p_{w^*})\,\delta^{3}(\vec{p}_{w^*})\right] \\
&\phantom{xxxxxxxxxxxxxxxxxxx}
=  \int \prod_{e^*_t} dg_{e^*_t} \prod_{e^*_{e,f}}dg_{e^*_{e,f}} \prod_{w^*}\; \left[ -\frac{1}{8} \eta^\square_{w^*}\, \eta_{w^*} \left(\cos\theta_{w^*} + 1\right)\right]\delta^3\left(\sin\theta_{w^*}\, \vec{n}_{w^*}\right),
\end{split}\label{quant7}
\ee
where $\sin\theta\,\, \vec{n}$ is the vector parameterizing the $\SU(2)$ subgroup.   By inspection, we find that the above integrand is peaked on $G_{w^*} = \id$, with the correct numerical factor.  This is in marked contrast with the $\SU(2)$ case, where one must insert an appropriate $G_{w^*}$-dependent observable to kill a second peak for which the $\SU(2)$ part of the holonomy is equal to $-\id$ instead of $+\id$.\footnotemark
\footnotetext{In fact, even here, the delta function over the even sector of $\UOSP(1|2)$ has two peaks.  The other peak is:
\begin{equation*}
\tilde{\id}  =\left(
\begin{array}{ccc}
1 & 0 & 0\\
0 & -1 & 0\\
0 & 0 & -1
\end{array}\right).
\end{equation*}
Fortuitously, the delta-function over the odd sector contains a factor $(\cos\theta + 1)$ which kills the second peak. Such a factor was introduced by hand for the $\SU(2)$ Ponzano-Regge model in \cite{Bobs} in order to kill this same second peak.}
The fact that the path integral of the discrete supersymmetric action automatically kills the second peak in the group element is a noticeable improvement on the standard $\SU(2)$ discrete path integral which leads to a $\delta$-function over $\SO(3)$ thus peaked on both $+\id$ and $-\id$ from the $\SU(2)$ viewpoint \cite{PR1}. This could only be resolved by adding suitable measure factors in the path integral in order to remove this second peak by hand (e.g. see  \cite{Bobs}). Here, in our supersymmetric framework, it is the presence of matter itself that takes care of this issue. Intuitively, the equation of motion \Ref{class8} imposes that the fermionic field $\psi$ has a trivial parallel transport, and thus distinguishes between $+\id$ and $-\id$ holonomies since we are dealing with spinors. Maybe such a feature can be generalised beyond the supersymmetric theory and we could conjecture that the $\SU(2)$ holonomy is necessarily peaked on $+\id$ whenever fermions are present in the theory.
The partition function may be written as:
\be
\cZ_{\Delta, \UOSP(1|2)} =  \int \prod_{e^*_t} dg_{e^*_t} \prod_{e^*_{e,f}}dg_{e^*_{e,f}} \prod_{w^*}\:\delta(G_{w^*}).\label{quant8}
\ee
As an aside, we note that we integrate with respect to the $g_{e^*_{e,f}}$ variables, we in essence glue the wedges within a face together and the amplitude becomes:
\be
\cZ_{\Delta, \UOSP(1|2)} =  \int \prod_{e^*} dg_{e^*}  \prod_{f^*}\:\delta(G_{f^*}).\label{quant8a}
\ee
For $\UOSP(1|2)$, there also exists a Plancherel measure, with respect to which the $\delta$-function may be decomposed as:
\be
\delta(G) = \sum_{j\in\frac{1}{2}\mathbb{N}} \Sdim_j\; \Schi^j(G),\label{quant9}
\ee
where $\Sdim_j = (-1)^{2j}$ is the superdimension, and $\Schi^j(G)$ is the supercharacter, that is, the {\it supertrace} of the representation matrix ${}^jT^{(k\,m)}{}_{(l\,n)}(G)$ (see Appendix \ref{osp} for details).

Having performed the Peter-Weyl decomposition on each $\delta$-function, we are left to manipulate the representation functions and integrate with respect to the group variables.  This is necessarily an arduous task in the supergroup case since permuting matrix elements may introduce factors of $(-1)$ if the matrix element is an element of the even or odd sector of the Grassmannian algebra.  Fortunately, there exists an efficient method to manage the bookkeeping of such factors.  This is a graphical calculus prevalent in some approaches to braided monoidal categories, and presented for the supergroup case in \cite{livineoeckl}. We shall not review it in all its glory here, but shall give just a bare bones description.

First of all, let us draw the template upon which we shall illustrate a sample of all our subsequent amplitudes.  Consider a triangle in the simplicial complex along with the faces dual to its three edges.  Moreover, consider the wedges which constitute each of these faces:
\begin{figure}[H]\centering
\begin{minipage}{0.4\linewidth}\centering
\includegraphics[width =7cm]{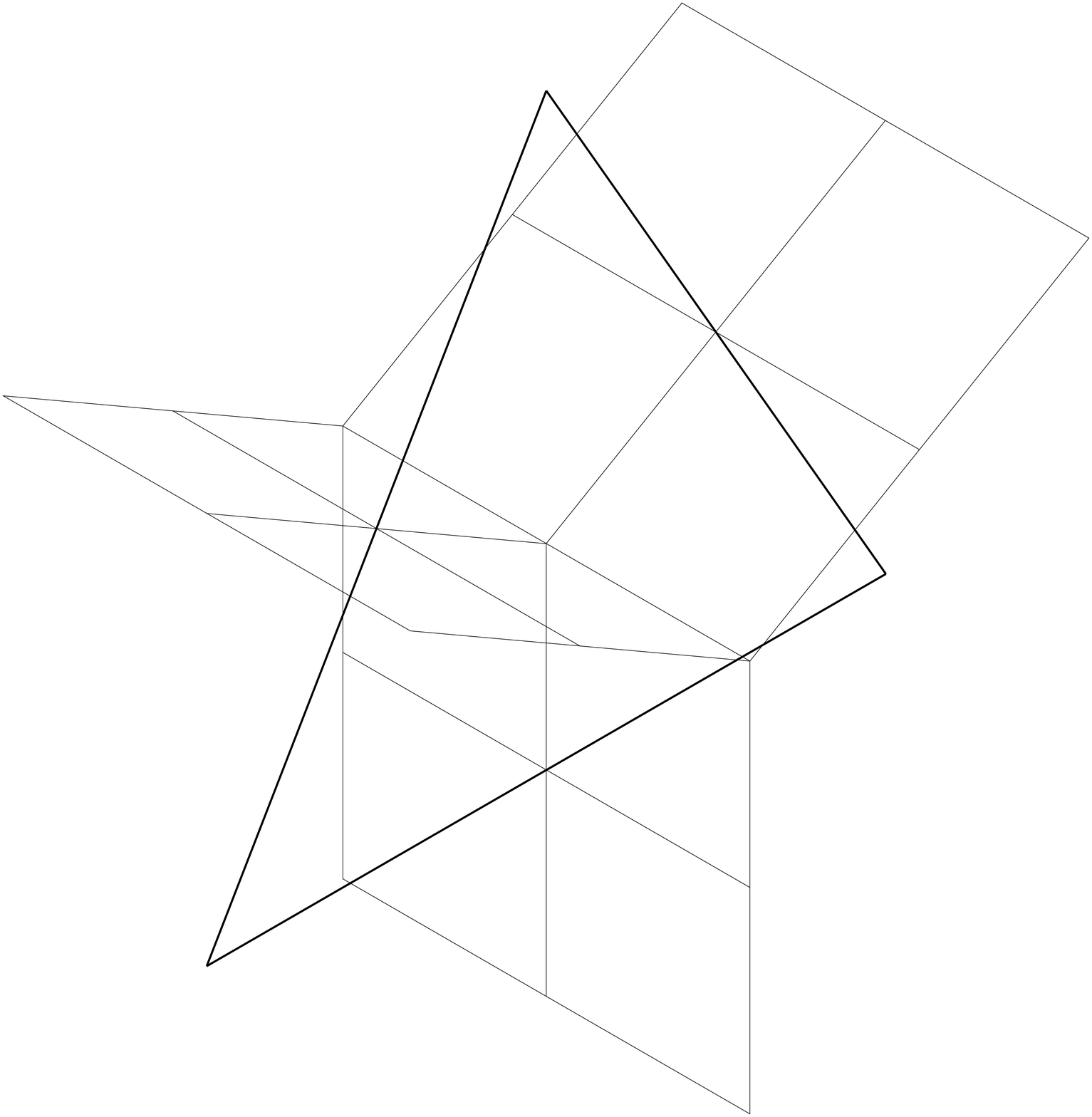}
\end{minipage}
\begin{minipage}{0.4\linewidth}\centering
\psfrag{h}{$g_{e^*_{e,f}}$}
\psfrag{g}{$g_{e^*_t}$}
\psfrag{j}{$j_{w^*}$}
\includegraphics[width =7cm]{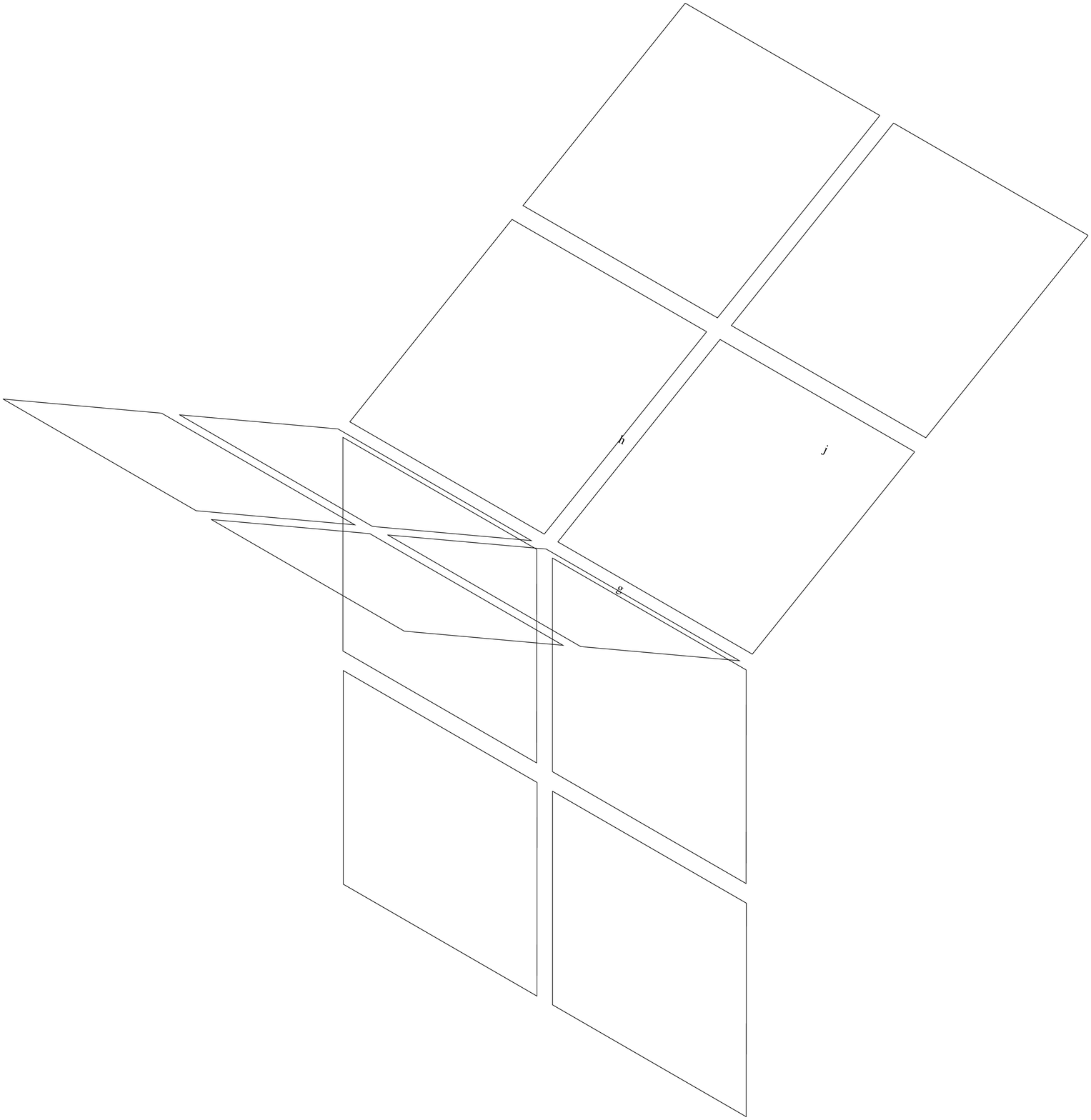}
\end{minipage}
\caption{\label{sample}Sample of the labelling of the simplicial complex and its topological dual.}
\end{figure}
As we see from \eqref{quant8},  the edges of the wedge are labelled by group elements $g_{e^*_t}$ and $g_{e^*_{e,f}}$.  Also, a representation $j_{w^*}$ labels each wedge.  As for the amplitude itself, to each supercharacter in \eqref{quant8}, we note that there is a loop labelled with the representation $j_{w^*}$ and the holonomy around the wedge. (Since the factor $\Sdim_{j_{w^*}}$ is actually the supercharacter of the identity element, we should include another loop with trivial argument for each wedge, but to simplify the illustration, we include this implicitly.) Importantly, we can see that several wedges may share the same group element.  The ultimate power of this formalism is that one can manipulate the diagrams using the rule:
\begin{figure}[H]\centering
\begin{minipage}{0.4\linewidth}\centering
\psfrag{s1}{${}^{(l_1n_1)}$}
\psfrag{t1}{${}^{(k_1m_1)}$}
\psfrag{s2}{${}^{(l_rn_r)}$}
\psfrag{t2}{${}^{(k_rm_r)}$}
\psfrag{j1}{${}^{j_1}$}
\psfrag{jr}{${}^{j_r}$}
\psfrag{g}{$g$}
\includegraphics[width =4cm]{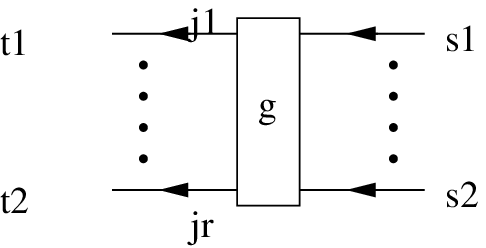}
\end{minipage}
\begin{minipage}{0.1\linewidth}\centering$ = $\end{minipage}
\begin{minipage}{0.4\linewidth}\centering
\psfrag{s1}{${}^{(l_1n_1)}$}
\psfrag{t1}{${}^{(k_1m_1)}$}
\psfrag{s2}{${}^{(l_rn_r)}$}
\psfrag{t2}{${}^{(k_rm_r)}$}
\psfrag{j1}{${}^{j_1}$}
\psfrag{jr}{${}^{j_r}$}
\psfrag{g}{$g$}
\includegraphics[width = 4cm]{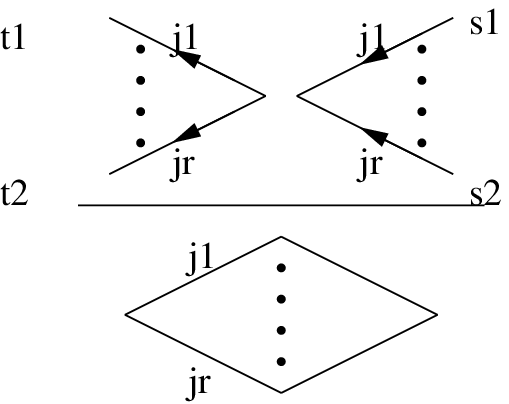}
\end{minipage}
\caption{\label{rule}Diagrammatic rule to convert representation functions to intertwiners.}
\end{figure}
where:
\begin{figure}[H]\centering
\begin{minipage}{0.4\linewidth}\centering
$I^{j_1}_{(k_1m_1)}{}^{\dots}_{\dots}\;{}^{j_r}_{(k_rm_r)}    $
\end{minipage}
\begin{minipage}{0.1\linewidth}\centering$=$\end{minipage}
\begin{minipage}{0.4\linewidth}\centering
\psfrag{a1}{${}^{j_1}$}
\psfrag{a3}{${}^{j_r}$}
\psfrag{m1}{${}^{(k_1\,m_1)}$}
\psfrag{m3}{${}^{(k_r\,m_r)}$}
\includegraphics[width = 2.7cm]{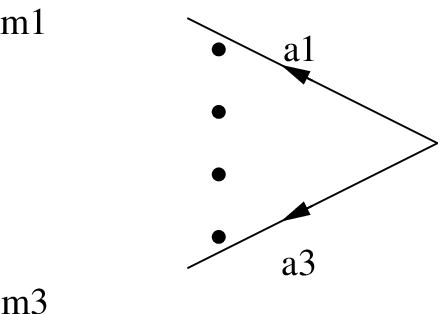}
\end{minipage}
\caption{\label{interrupt} Graphical representation of intertwiner.}
\end{figure}
\noindent $I$ is an intertwiner on the representation space of $\UOSP(1|2)$.\footnote{There is a computational subtlety in producing the above rule explicitly in terms of objects in the representation theory of $\UOSP(1|2)$, which we describe in Appendix \ref{theta}. }  Applying this rule everywhere possible we find that representations attached to the wedges within a given face $f^*$ are forced to coincide leaving just one which we shall denote $j_e = j_{f^*}$.  On top of this,  the diagram factorises as follows:
\begin{figure}[H]\centering
\psfrag{Z}{$\cZ_{\Delta, \UOSP(1|2)} $}
\psfrag{a}{$\displaystyle=\sum_{\{j\}}$}
\psfrag{b}{$\displaystyle\prod_{f^*}$}
\psfrag{c}{$\displaystyle\prod_{e^*}$}
\psfrag{d}{$\displaystyle\prod_{v^*}$}
\psfrag{1}{\scriptsize{-1}}
\includegraphics[width = 9cm]{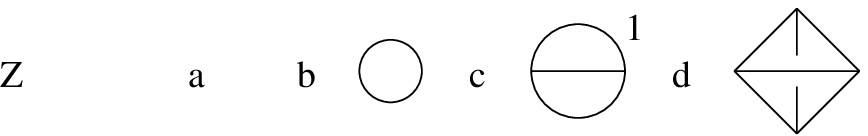}
\end{figure}
Thus, it is just a matter of evaluating these subdiagrams.  We do this explicitly in Appendix \ref{diagram} and to conclude, the amplitude is roughly a product of supersymmetric $\{6j\}$-symbols and takes the following form:
\be\label{quant10}
\cZ_{\Delta, \UOSP(1|2)}
 \displaystyle=\sum_{\{j\}} \prod_e(-1)^{2j_e} \prod_{f} (-1)^{\lfloor j_{e_1}+ j_{e_2} + j_{e_3}\rfloor_f}\prod_t \left[
\begin{array}{ccc}
j_{e_1} & j_{e_2} & j_{e_3}\vspace{0.1cm}\\
j_{e_4} & j_{e_5} & j_{e_6}
\end{array}
\right]_t \;.
\ee

\section{Analysis of N=1 supersymmetric BF theory}
\label{susyone}

We continue the investigation of $N=1$ supersymmetric spin foams, initiated in \cite{livineoeckl}.      In Section \ref{derivation}, we derived the $\UOSP(1|2)$ spin foam model directly from a discrete path integral.  Ultimately, the amplitude takes the form given in  \eqref{quant10}.  Considering the amplitudes of the sub-simplices of $\Delta$ individually, the edges, triangles and tetrahedra carry the weights: $(-1)^{2j_e}$ (the superdimension); $(-1)^{\lfloor j_{e_1}+j_{e_2}+j_{e_3}\rfloor_f}$ (the normalisation of the supersymmetric $\{3j\}$-symbol); and the supersymmetric $\{6j\}$-symbol, respectively.  Moreover, given a fixed triangulation $\Delta$, the sum ultimately includes all admissible configurations of irreducible $\UOSP(1|2)$ representations attached to the edges of the triangulation.  These configurations are labelled $\{j_e\}$ ($j_e\in\f{\N}{2}$).   To be an admissible configuration,  the representations must satisfy triangle inequalities {\it but} the familiar closure condition is relaxed.  That is to say, if $e_1,e_2,e_3\in \partial f$, then:
\be\label{susy1}
\begin{array}{rcl}
&&|j_a - j_b| \leq j_c \leq j_a+j_b, \quad \textrm{where $a,b,c$ are a permutation of $e_1,e_2,e_3$.}\\
but &&\\
&&j_{e_1}+j_{e_2}+j_{e_3}\in\N, \quad \textrm{or}\quad j_{e_1}+j_{e_2}+j_{e_3}\in\N+\f{1}{2}.
\end{array}
\ee
The spin foam amplitude is a function of these representations, as is conventional.

From the classical standpoint, we derived this quantum amplitude beginning with an action displaying supersymmetric gauge invariance.  But since the gauge group $\UOSP(1|2)$ is built upon the familiar $\SU(2)$ Lie group, there is a nice $\SU(2)$ structure nested inside this overarching supersymmetric one.  As we uncovered in Section \ref{derivation}, upon making this $\SU(2)$ structure explicit, one sees that this theory is one describing gravity coupled to Grassmann-valued spin-$\frac{1}{2}$ fields. Now that we have the quantum amplitude in a well-defined discrete setting, we expect that within the supersymmetric partition function lie amplitudes pertaining to gravity coupled to these spin-$\frac{1}{2}$ fermionic fields.   Indeed, our main aim in this paper is to clarify this correspondence.

Perhaps some intuition for what might happen can be gained by examining a generic parallel transport matrix in a given representation $j_e$, namely ${}^{j_e}T^{(k_em_e)}{}_{(l_en_e)}(g_{e^*})$.  Since each tetrahedron contains its own frame of reference, this matrix describes the change in certain properties pertaining to the edge $e\in\Delta$ as one moves from one tetrahedron to the next along $e^* \in \Delta^*$ .  One such property is the length of the edge, $e$, as seen in each tetrahedron, which is given by the $\SU(2)$ sub-module (labelled by $k_e,\, l_e$) pertaining each tetrahedron.  To spell it out, ${}^{j_e}T^{(k_em_e)}{}_{(l_en_e)}(g_{e^*}) : V^{l_e} \rightarrow V^{k_e}$, so that the length of the edge as viewed from the initial tetrahedron is $l_e+\frac{1}{2}$, while from the final tetrahedron is seems that the edge has length $k_e+\frac{1}{2}$.\footnote{In pure gravity, the length an edge of the spin foam is given by $\displaystyle\frac{\dim_{k_e}}{2} = k_e + \frac{1}{2}$.}
%
At a hand-waving level, the fact that different observers (here tetrahedra) see different lengths for the same edge comes from a non-vanishing torsion in the theory.
Since both $k_e, l_e$ can take the values $j_e, j_e - \frac{1}{2}$ freely, the length of the edge may change from reference frame to reference frame. Note, however, that the matrix elements fall into two classes.  On the one hand there are the cases where the edge length does not change: $\displaystyle{}^{j_e}T^{(j_e\, m_e)}{}_{(j_e\,n_e)}(g_{e^*})$ and $\displaystyle{}^{j_e}T^{(j_e-\frac{1}{2}\,m_e)}{}_{(j_e-\frac{1}{2}\,n_e)}(g_{e^*})$.  For example, let us examine:
\be\label{susy2}
{}^{j_e}T^{(j_e\,m_e)}{}_{(j_e\, n_e)} (g_{e^*})
				=(1-\frac{1}{4}j_e\,\eta_{e^*}^\square\eta_{e^*})\; {}^{j_e}\!D^{m_e}{}_{n_e}(\Omega_{e^*}),
\ee
where  $\Omega = \{\psi,\, \theta,\, \phi\}$.  Each element is even in the Grassmann algebra.  We propose that the $\cO((\eta_{e^*}^\square\eta_{e^*})^0)$ term corresponds to no-fermion propagation, while the $\cO((\eta_{e^*}^\square\eta_{e^*})^1)$ term corresponds to the propagation of both a fermion and an anti-fermion, which together yield a bosonic contribution.  On the other hand, there are the cases where the edge length does change: $\displaystyle{}^{j_e}T^{(j_e\,m_e)}{}_{(j_e-\frac{1}{2}\,n_e)}(g_{e^*})$ and $\displaystyle{}^{j_e}T^{(j_e-\frac{1}{2}\,m_e)}{}_{(j_e,n_e)}(g_{e^*})$. this time, let us examine:
\be\label{susy3}
\begin{split}
&{}^{j_e}T^{(j_e\, m_e)}{}_{(j_e-\frac{1}{2}\, n_e)} (g_{e^*})\\
&\phantom{xxxxxxxx}
				=\left[-\frac{1}{2}\sqrt{j_e+n_e+\frac{1}{2}}\;\eta_{e^*}^\square\; {}^{j_e}\!D^{m_e}{}_{n_e+\frac{1}{2}}(\Omega_{e^*}) - \frac{1}{2}\sqrt{j_e-n_e+\frac{1}{2}}\;\eta_{e^*}\; {}^{j_e}\!D^{m_e}{}_{n_e-\frac{1}{2}}(\Omega_{e^*})\right].
\end{split}
\ee
Note that each term is linear in an odd Grassmann variable, so we shall interpret this as the propagation of a fermion or an anti-fermion.  Albeit a somewhat loose description of what is happening, it is essentially correct and once we describe this in terms of an $\SU(2)$ lattice gauge theory coupled to Grassmann fields, we shall see this prescription become more precise.

Let us begin to analyze the quantum amplitudes directly. We shall first make the embedded $\SU(2)$ substructure explicit at the level of representations by expanding the $\UOSP(1|2)$ $\{6j\}$-symbols in terms of $\SU(2)$ $\{6j\}$-symbols (\ref{susy6j}):
\be
\begin{array}{rcl}
\left[\begin{array}{ccc}
j_1 & j_2 & j_3\vspace{0.1cm}\\
j_4 & j_5 & j_6
\end{array}\right]

 & = & \displaystyle\sum_{\substack{k_i\\1\leq i\leq 6}}(-1)^{\sum_{a=1}^62(j_a-k_a)(\lambda_a+ 1)}
  B^{j_1j_2j_3}_{k_1k_2k_3}\; B^{j_5j_6j_1}_{k_5k_6k_1}\; B^{j_6j_4j_2}_{k_6k_4k_2}\; B^{j_4j_5j_3}_{k_4k_5k_3}\;
\left\{\begin{array}{ccc}
k_1 & k_2 & k_3\vspace{0.1cm}\\
k_4 & k_5 & k_6
\end{array}\right\},
\end{array}\nonumber
\ee
where we choose the parity $\lambda_a=2j_a$ and we sum over $k_i=j_i$ or $=j_i-\f12$.
Since each triangle is shared by two tetrahedra, we may repartition the amplitude as follows:
\be\label{susy4}
\begin{array}{rcl}
\cZ_{\Delta,\UOSP(1|2)} &=& \displaystyle\sum_{\{j, k\}}  \prod_{e} (-1)^{2j_e}
\prod_f \left[    (-1)^{ \lfloor j_{e_1}+ j_{e_2} + j_{e_3} \rfloor }  B^{j_{e_1}j_{e_2}j_{e_3}}_{k_{e_1}k_{e_2}k_{e_3}}  B^{j_{e_1}j_{e_2}j_{e_3}}_{k'_{e_1}k'_{e_2}k'_{e_3}} \right] _f

\\

&& \phantom{xxxxxxxxxxx}\displaystyle \times
\prod_{t}\left[ (-1)^{\sum_{a =1}^6 2(j_{e_a} - k_{e_a} )(2j_{e_a}+1)}\left\{
\begin{array}{ccc}
k_{e_1} & k_{e_2} & k_{e_3}\vspace{0.1cm}\\
k_{e_4} & k_{e_5} & k_{e_6}
\end{array}
\right\}\right]_t

\end{array},
\ee
where the appropriate definitions are given in Appendix \ref{intertwiners}. Let us insist that although we sum over one label $j_e$ per edge, we are also summing over one label $k_{e,t}$ per edge $e$ and per tetrahedron $t\ni e$ to which the edge belongs.

For future reference, let us scrutinise the triangle amplitude:
\be\label{susy5}
A_f^{\{j\}}((k_{e_1},k_{e_2},k_{e_3}; k'_{e_1}, k'_{e_2}, k'_{e_3}) =  \left[(-1)^{ \lfloor j_{e_1}+ j_{e_2} + j_{e_3} \rfloor }  B^{j_{e_1}j_{e_2}j_{e_3}}_{k_{e_1}k_{e_2}k_{e_3}}  B^{j_{e_1}j_{e_2}j_{e_3}}_{k'_{e_1}k'_{e_2}k'_{e_3}} \right] _f
\ee
  There are a number of forms this can take depending on the values of $k_e$ and $k_e'$.  But there are certain basic forms that they follow.  To save space, let us denote the element $k = j$ by $\uparrow$ and $k = j-\frac{1}{2}$ by $\downarrow$.  Thus, $A^{\{j\}}_f(j_{e_1},j_{e_2},j_{e_3}; j_{e_1}\!-\frac{1}{2}, j_{e_2}\!-\frac{1}{2}, j_{e_3}) = A^{\{j\}}_f(\uparrow,\uparrow, \uparrow ; \downarrow,\downarrow, \uparrow)$.  We note also that the amplitudes are symmetric with respect to the interchange of $\{k_1,k_2,k_3\}$ and $\{k'_1, k'_2, k'_3\}$.  The possible configurations are (up to flipping entirely $\{k_1,k_2,k_3\}$ with $\{k'_1, k'_2, k'_3\}$):
\be\label{amps1}
\begin{array}{l}
j_{e_1}+j_{e_2}+j_{e_3} \in \mathbb{N} \\
 \\
{\setlength{\extrarowheight}{0.3cm}
\begin{array}{rcl}
A_f^{\{j\}} (\ua,\ua,\ua; \ua,\ua,\ua ) & = &   (-1)^{J} (j_{e_1} + j_{e_2} + j_{e_3} + 1)\\
A_f^{\{j\}} (\ua,\da,\da; \ua,\da,\da ) & = &	(-1)^{J} (-j_{e_1} + j_{e_2} + j_{e_3})\\
A_f^{\{j\}} (\da,\ua,\da; \da,\ua,\da ) & = &	(-1)^{J} (j_{e_1} - j_{e_2} + j_{e_3})\\
A_f^{\{j\}} (\da,\da,\ua; \da,\da,\ua ) & = &	(-1)^{J} (j_{e_1} + j_{e_2} - j_{e_3})\\
A_f^{\{j\}} (\ua,\ua,\ua; \ua,\da,\da ) & = &	(-1)^{J + (2j_{e_1}+1)} \sqrt{(j_{e_1} + j_{e_2} + j_{e_3} + 1) (-j_{e_1} + j_{e_2} + j_{e_3})}\\
A_f^{\{j\}} (\ua,\ua,\ua; \da,\ua,\da ) & = &	(-1)^{J + (2j_{e_2}+1)} \sqrt{(j_{e_1} + j_{e_2} + j_{e_3} + 1) (j_{e_1} - j_{e_2} + j_{e_3})}\\
A_f^{\{j\}} (\ua,\ua,\ua; \da,\da,\ua ) & = &	(-1)^{J + (2j_{e_3}+1)} \sqrt{(j_{e_1} + j_{e_2} + j_{e_3} + 1) (j_{e_1} + j_{e_2} - j_{e_3})}\\
A_f^{\{j\}} (\ua,\da,\da; \da,\ua,\da ) & = &	(-1)^{J + (2j_{e_1}+1)+(2j_{e_2}+1)} \sqrt{(-j_{e_1} + j_{e_2} + j_{e_3} ) (j_{e_1} - j_{e_2} + j_{e_3})} \\
A_f^{\{j\}} (\ua,\da,\da; \da,\da,\ua ) & = &	(-1)^{J + (2j_{e_3}+1)+(2j_{e_1}+1)} \sqrt{(-j_{e_1} + j_{e_2} + j_{e_3}) (j_{e_1} + j_{e_2} - j_{e_3})}\\
A_f^{\{j\}} (\da,\ua,\da; \da,\da,\ua ) & = &	(-1)^{J + (2j_{e_2}+1)+(2j_{e_3}+1)} \sqrt{(j_{e_1} - j_{e_2} + j_{e_3}) (j_{e_1} + j_{e_2} - j_{e_3})}\phantom{xxxxxxxxxxxxx}
\end{array}}
\end{array}
\ee
and
\be\label{amps2}
\begin{array}{l}

 j_{e_1}+j_{e_2}+j_{e_3} \in \mathbb{N} +\frac{1}{2}\\
 \\
{\setlength{\extrarowheight}{0.3cm}
\begin{array}{rcl}
A_f^{\{j\}} (\da,\da,\da; \da,\da,\da )  & = &  (-1)^{J-\frac{1}{2}} (j_{e_1} + j_{e_2} + j_{e_3} + \frac{1}{2})\\
A_f^{\{j\}} (\da,\ua,\ua; \da,\ua,\ua )  & = &  (-1)^{J-\frac{1}{2}} (-j_{e_1} + j_{e_2} + j_{e_3} + \frac{1}{2})\\
A_f^{\{j\}} (\ua,\da,\ua; \ua,\da,\ua )  & = &  (-1)^{J-\frac{1}{2}} (j_{e_1} - j_{e_2} + j_{e_3} + \frac{1}{2}) \\
A_f^{\{j\}} (\ua,\ua,\da; \ua,\ua,\da )  & = &  (-1)^{J-\frac{1}{2}}(j_{e_1} + j_{e_2} - j_{e_3} + \frac{1}{2})\\
A_f^{\{j\}} (\da,\da,\da; \da,\ua,\ua )  & = & (-1)^{J-\frac{1}{2} + (2j_{e_1}+1)} \sqrt{(j_{e_1} + j_{e_2} + j_{e_3}  + \frac{1}{2}) (-j_{e_1} + j_{e_2} + j_{e_3} + \frac{1}{2})}\\
A_f^{\{j\}} (\da,\da,\da; \ua,\da,\ua )  & = &  (-1)^{J-\frac{1}{2} + (2j_{e_2}+1)} \sqrt{(j_{e_1} + j_{e_2} + j_{e_3}  + \frac{1}{2}) (j_{e_1} - j_{e_2} + j_{e_3} + \frac{1}{2})}\\
A_f^{\{j\}} (\da,\da,\da; \ua,\ua,\da )  & = &  (-1)^{J-\frac{1}{2} + (2j_{e_3}+1)} \sqrt{(j_{e_1} + j_{e_2} + j_{e_3} + \frac{1}{2}) (j_{e_1} + j_{e_2} - j_{e_3} + \frac{1}{2})}\\
A_f^{\{j\}} (\da,\ua,\ua; \ua,\da,\ua )  & = &  (-1)^{J-\frac{1}{2} + (2j_{e_1}+1)+(2j_{e_2}+1)} \sqrt{(-j_{e_1} + j_{e_2} + j_{e_3} + \frac{1}{2} ) (j_{e_1} - j_{e_2} + j_{e_3} + \frac{1}{2})} \\
A_f^{\{j\}} (\da,\ua,\ua; \ua,\ua,\da )  & = & (-1)^{J-\frac{1}{2} + (2j_{e_3}+1)+(2j_{e_1}+1)} \sqrt{(-j_{e_1} + j_{e_2} + j_{e_3} + \frac{1}{2}) (j_{e_1} + j_{e_2} - j_{e_3} + \frac{1}{2})}\\
A_f^{\{j\}} (\da,\da,\ua; \ua,\ua,\da )  & = &  	(-1)^{J-\frac{1}{2} + (2j_{e_2}+1)+(2j_{e_3}+1)} \sqrt{(j_{e_1} - j_{e_2} + j_{e_3} + \frac{1}{2}) (j_{e_1} + j_{e_2} - j_{e_3} + \frac{1}{2})}\phantom{xxxxx}
\end{array}}
\end{array}
\ee
where $J = j_{e_1} + j_{e_2} + j_{e_3}$. There are 32 configurations in total (20 are shown here and the other 12 are obtained by utilizing the symmetry given above).  These configurations split into two subsets depending on whether $J\in \mathbb{N}$ or $J\in \mathbb{N}+\frac{1}{2}$.  Thus, 16 are admissible at any instance.    Following on from what we mentioned just a little earlier, if the first three arrows do not differ from the second three, then the amplitude contributes to the even or \lq bosonic' sector of the theory,  while if there are two flips, then the amplitude contributes to the fermionic sector.
Remember that there can not be one or three flips due to the parity condition for the existence of $\SU(2)$ intertwiners.
Thus, the top four listed in each subset above are \lq bosonic' amplitudes while the rest are \lq fermionic' amplitudes.

Let us examine a bosonic amplitude:
\be\label{susy6}
A_f^{\{j\}} (\ua,\ua,\ua; \ua,\ua,\ua )  =    (-1)^{J} (j_{e_1} + j_{e_2} + j_{e_3} + 1).
\ee
The amplitude comes about from the coupling of $\displaystyle{}^{j_{e_1}}T^{(j_{e_1}m_{e_1})}{}_{(j_{e_1}n_{e_1})}(g_{e^*})$, $\displaystyle{}^{j_{e_2}}T^{(j_{e_2}m_{e_2})}{}_{(j_{e_2}n_{e_2})}(g_{e^*})$ and $\displaystyle{}^{j_{e_3}}T^{(j_{e_3}m_{e_3})}{}_{(j_{e_3}n_{e_3})}(g_{e^*})$.  Therefore, as we can see from \eqref{susy2}, in the product of these three factors,  the coefficient of  $(\eta^\square_{e^*}\eta_{e^*})$ is $(j_{e_1} + j_{e_2} + j_{e_3} + 1)$ (remembering that there is a $(\eta_{e^*}^\square\eta_{e^*})$ term in the measure). In fact, we interpret that this coefficient gets contributions from four different sources.  There may be no-fermion propagation {\it or} fermion - anti-fermion propagation on the edge $e_1$ {\it or} the edge $e_2$ {\it or} the edge $e_3$ (which contribute the $1$, $j_{e_1}$, $j_{e_2}$, $j_{e_3}$ pieces, respectively).

Now for a fermionic amplitude:
\be\label{susy7}
A_f^{\{j\}} (\ua,\ua,\ua; \ua,\da,\da )  = 	(-1)^{J-\frac{1}{2} + (2j_{e_1}+1)} \sqrt{(j_{e_1} + j_{e_2} + j_{e_3} + 1) (-j_{e_1} + j_{e_2} + j_{e_3})}.\\
\ee
Once again we see that this arises from coupling $\displaystyle{}^{j_{e_1}}T^{(j_{e_1}m_{e_1})}{}_{(j_{e_1}n_{e_1})}(g_{e^*})$, $\displaystyle{}^{j_{e_2}}T^{(j_{e_2}m_{e_2})}{}_{(j_{e_2}-\frac{1}{2}n_{e_2})}(g_{e^*})$ and $\displaystyle{}^{j_{e_3}}T^{(j_{e_3}m_{e_3})}{}_{(j_{e_3}-\frac{1}{2}n_{e_3})}(g_{e^*})$.
It is not possible to see how the coefficient of $\eta^\square_{e^*}\eta_{e^*}$ comes about directly, but for our interpretational purpose here, the coefficient gets a contribution from a fermion or anti-fermion propagation on both edges $e_2$ and $e_3$ while there is no-fermion propagation on edge $e_1$.

To conclude this section, while all the tetrahedra sharing an edge are colored with the same $\UOSP(1|2)$ representation $j_e$, they need not necessarily share the same $\SU(2)$ label $k_{e}$.   This is to be expected since the $\SU(2)$-modules lie within larger $\UOSP(1|2)$-modules.  Heuristically, we can divide the triangle amplitudes into two classes, differentiated by the condition  $k_{e_a} = k'_{e_a}$  for each edge of the triangle (keeping in mind that the triangle belongs to two tetrahedra).  Those which satisfy this condition are \lq bosonic', while those triangles for which this condition is not satisfied are \lq fermionic'.  We shall now make this interpretation precise, by Fourier transforming to the space of functions on $\SU(2)$.  It is rather difficult to do this succinctly, but we shall circumvent some of the clumsiness by jumping to the other side of the computation and working back.  Indeed, it is more instructive to do so.

\subsection{$\SU(2)$ spin foam amplitudes}
\label{sutwo}
The $\SU(2)$ Ponzano-Regge state sum amplitudes have the same fundamental building blocks as their $\UOSP(1|2)$ counterparts.   In terms of lattice gauge theory variables, one can write the amplitude as in \eqref{quant8a}:
\be\label{sutwo0}
\cZ_{\Delta, \SU(2)} = \int \prod_{e^*}dg_{e^*} \prod_{f^*} \delta(g_{f^*}).
\ee
After integration, we once again see a diagram like Fig.\ref{sample} but this time it is labelled by irreducible representations and group elements of $\SU(2)$.  As usual, after manipulation using an analogous rule to Fig.\ref{rule}, the amplitude factorises and we arrive at:

\begin{figure}[H]\centering
\psfrag{Z}{$\cZ_{\Delta, \SU(2)} $}
\psfrag{a}{$\displaystyle=\sum_{\{k\}}$}
\psfrag{b}{$\displaystyle\prod_{f^*}$}
\psfrag{c}{$\displaystyle\prod_{e^*}$}
\psfrag{d}{$\displaystyle\prod_{v^*}$}
\psfrag{1}{\scriptsize{-1}}
\includegraphics[width = 9cm]{graphamp.eps}
\end{figure}

\noindent Upon evaluating the corresponding diagrams, one gets the familiar Ponzano-Regge ampitude:
\be
\label{sutwo1}
\cZ_{\Delta, \SU(2)} = \sum_{\{k\}} \prod_e (-1)^{2k_e}(2k_e+1)  \prod_{f} (-1)^{ (k_{e_1}+ k_{e_2} + k_{e_3})_f}\prod_t
\left\{
\begin{array}{ccc}
k_{e_1} & k_{e_2} & k_{e_3}\vspace{0.1cm}\\
k_{e_4} & k_{e_5} & k_{e_6}
\end{array}
\right\}_t \; ,
\ee
To make a connection with supersymmetric amplitudes,  we must first rewrite this amplitude more appropriately to the supersymmetric context.  As pointed out in \cite{livineoeckl}, one has some freedom in the properties of the representations occurring in the decomposition of functions over the group.  For a start, one may endow the representations with a $\mathbb{Z}_2$-grading, so that vectors in $V^{k_e}$, $k_e\in\mathbb{N}+\frac{1}{2}$ are odd and vectors in $V^{k_e}$, $k_e\in\mathbb{N}$ are even.  Moreover, one can choose the inner product on an irreducible representation to be either positive definite or negative definite. None of these possibilities affects the decomposition of the $\delta$-function.  Let us denote the usual characters by $\chi$ and the grades ones by ${\chi}^{\pm}$ where $\pm$ labels the choice of inner product, then:
\be\dim_k \; {\chi}^{k}(g) = \dim_{k,\pm} \; {\chi}^{k,\pm}(g)\ee
where $\dim_{k,\pm} := {\chi}^{k,\pm}(\id)$.  So, we may write:
\be
\begin{split}
\delta(g) = \sum_{k} \dim_{k}\;{\chi}^{k}(g) = \sum_{k} \dim_{k,\pm}\; {\chi}^{k,\pm}(g) = &
\frac{1}{2}\sum_{j\in \mathbb{N}+\frac{1}{2}} \left( \dim_{j,+}\; {\chi}^{j,+}(g) + \dim_{j-\frac{1}{2}, + } \; {\chi}^{j-\frac{1}{2}, + }(g)\right)\\
&\phantom{xx}+\frac{1}{2}\sum_{j\in \mathbb{N}} \left( \dim_{j,+}\; {\chi}^{j,+}(g) + \dim_{j-\frac{1}{2}, - } \; {\chi}^{j-\frac{1}{2}, - }(g)\right).
\end{split}
\ee
Thus, instead of viewing the decomposition as a sum over irreducible representations $V^{k}$, one can view it as decomposed over the representations $R^{j} = V^{j} \oplus V^{j - \frac{1}{2}}$, where the representations are graded, and the inner product on $V^{j}\subset R^{j}$ is positive for all $j$, while that on $V^{j-\frac{1}{2}} \subset R^{j}$ is positive for $j\in \mathbb{N} + \frac{1}{2}$ and negative for $j\in\mathbb{N}$.  This choice is compatible the tensor product operation, and mimics the structure in the supersymmetric theory.   Thus, the $\SU(2)$ Ponzano-Regge spin foam amplitude may be recast as an amplitude depending on $R^{j}$.  Upon integration
of the group elements in \eqref{sutwo0}, we obtain the same diagrams as before, but the evaluation of the loop, theta, and tetrahedral diagrams depends on the grading and inner product, and we find that we can write the amplitude as:
\be\label{sutwo2}
\begin{array}{rcl}
\cZ_{\Delta, \SU(2)} &=& \displaystyle\sum_{\{j\}}\sum_{\{k\}} \prod_e (-1)^{2j_e}(2k_e+1)  \prod_{f} (-1)^{\sum_{a =1}^3 \left[2(j_{e_a} - k_{e_a} )(2j_{e_a}+1)+k_{e_a}\right]_f}

\\
&&\phantom{xxxxxxxxxxxx}\displaystyle\times
\prod_t  \left[(-1)^{\sum_{a =1}^6 2(j_{e_a} - k_{e_a} )(2j_{e_a}+1)}
\left\{
\begin{array}{ccc}
k_{e_1} & k_{e_2} & k_{e_3}\vspace{0.1cm}\\
k_{e_4} & k_{e_5} & k_{e_6}
\end{array}
\right\}\right]_t \; ,
\end{array}
\ee
We must stress, however, that although the amplitude looks different, it is merely a repartitioning of the original state-sum.  We evaluate the diagrams explicitly in Appendix \ref{diagram}.

Let us insist also on the fact that both the $j_e$'s and $k_e$'s depend only on the chosen edge: it is the same $k_e$ all around the corresponding plaquette and it does not change from tetrahedron to tetrahedron. The fluctuations of $k_e$ around the plaquette  that occur in the supersymmetric theory will come in when we insert fermions in the model, as explained below.

\subsection{Coupling matter: massless spinning fields}

Of course, looking at the $\SU(2)$ theory from the lattice gauge theory perspective, a group element cannot map between different irreducible representations.  In other words, the edge length cannot change as we move from tetrahedron to tetrahedron. Thus, the non-trivial matter is to allow for a change in edge length and this is where we expect the fermionic degrees of freedom to come into play.  We wish to insert fermionic observables into \eqref{sutwo2} so as to get contributions to $\cZ_{\Delta, \UOSP(1|2)}$.  This section will be concerned with the construction of these observables.

Noticing that the odd generators of $\UOSP(1|2)$ carry a spin-$\frac{1}{2}$ representation of $\SU(2)$, we follow that argument to its natural conclusion and trace a spin-$\frac{1}{2}$ representation through the spin foam. Furthermore, remembering the points expounded earlier, we expect that it should be embedded at the gluing point of two tetrahedra, and that if there is no change in edge length then there are either zero or two fermionic lines, while if the edge length changes then there is a single fermionic line embedded there.  Indeed, this is how things turn out in the end.

To begin, we need to have a wedge formulation in terms of holonomies and representations $R^j$.  One might expect that this is a trivial manipulation of \eqref{sutwo0}, but to maintain the $R^j$ structure, requires us to navigate certain subtleties.  Consider two adjacent wedges within a face, $w^*_1$ and $w^*_2$.  Each has an assigned reducible representation $R^{j_{w^*_1}}$ and $R^{j_{w^*_2}}$.  Whether a fermion observable is inserted or not, we wish wedges to couple only if ${j_{w^*_1}} = j_{w^*_2}$.
This is naturally the case for irreducible representations of $\SU(2)$, but not for reducible ones.  To see this, one need only notice that $V^{j_{w^*}}$ is contained in $R^{j_{w^*}}$ and $R^{j_{w^*}+\frac{1}{2}}$. Therefore, $R^{j_{w^*_1}}$ and $R^{j_{w^*_2}}$ will couple for $j_{w^*_2} = j_{w^*_1}, j_{w^*_1}\pm \frac{1}{2}$.  We can cure this ambiguity by inserting a projector into the holonomy matrix attached to each wedge.  It projects onto the highest weight state for $V^j\subset R^{j}$ and onto the lowest weight state for $V^{j-\frac{1}{2}}\subset R^{j}$.  
This projector is illustrated by:

\begin{figure}[H]\centering
\begin{minipage}{0.4\linewidth} \centering
\psfrag{j}{${}^{\pm, k_{w^*}}$}
\psfrag{s}{${}^{}$}
\psfrag{t}{${}^{\phantom{(k\,\,}}$}
\includegraphics[width = 4cm]{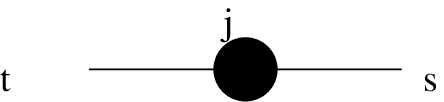}
\end{minipage}
\begin{minipage}{0.05\linewidth} \centering $=$\end{minipage}
\begin{minipage}{0.4\linewidth} \centering
$\displaystyle \left|k_{w^*}, \pm k_{w^*}\right>\quad (-1)^{2(j_{w^*}-k_{w^*})(2j_{w^*}+1)}\quad\left<k_{w^*}, \pm k_{w^*}\right|  $
\end{minipage}
\end{figure}
\noindent and it is inserted at the point $v^*_e$ in Fig.\ref{wedge}.

In effect, when one integrates over the $g_{e^*_{e,f}}$ variables, one gets a factor of $\delta_{j_{w^*_2}, j_{w^*_1}\pm \frac{1}{2}} = 0$ from the projectors.  Only the $j_{w^*_1} = j_{w^*_2}$ term survives.  Now,  we insert the fermionic observables $\cO_\cF$:
\be
Z_{\Delta, \SU(2), \cO_\cF} = \int \prod_{e^*_t} dg_{e^*_t} \prod_{e^*_{e,f}}dg_{e^*_{e,f}} \prod_{w^*} \sum_{\{j\}} A_{w^*}(g_{w^*}, j_{w^*})\; \cO_\cF(\{g\}, \{j\}).
\ee
So let us proceed to the definition of these observables.  Diagrammatically, we shall denote a segment of the fermionic observable by a dashed line. Furthermore, we shall need to introduce the projector onto the spin-up and spin-down states:
\begin{figure}[H]\centering
\begin{minipage}{0.4\linewidth} \centering
\psfrag{j}{$\pm$}
\psfrag{s}{}
\psfrag{t}{}
\psfrag{g}{}
\includegraphics[width = 4cm]{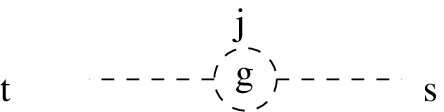}
\end{minipage}
\begin{minipage}{0.05\linewidth} \centering $=$\end{minipage}
\begin{minipage}{0.4\linewidth} \centering
$\displaystyle \left|\frac{1}{2}, \pm\frac{1}{2}\right>\quad\left<\frac{1}{2}, \pm\frac{1}{2}\right|  $
\end{minipage}
\end{figure}
\noindent  One would expect this projection to occur as part of the propagator for spin-$\frac{1}{2}$ fermions \cite{PR1, Weinberg}.  These projectors are inserted into the diagram once again at the points $v^*_e$ and are joined by parallel transport matrices, which closely follows the procedure for the insertion of matter observables in \cite{PR1}.   This charts the progress of the particle in the spin foam formulation.  We shall examine more clearly the geometric space-time interpretation of this fermionic path shortly.  The dashed line runs along between the wedges, since we want to allow for a change in edge length by $\frac{1}{2}$ as we move between tetrahedra.
Furthermore, the fermionic observable knows about the gravity sector.  We must insert an operator to extract various factors of edge-length: $\frac{\dim_k}{2}$. We shall denote this graphically by a clasp joining the fermion projector and the gravity projector:
\begin{figure}[H]\centering
\psfrag{j}{${}^{\pm, k_{w^*}}$}
\psfrag{f}{}
\includegraphics[width = 3cm]{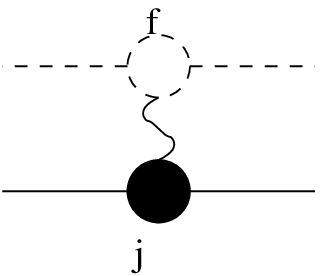}
\end{figure}
\noindent    We give some motivation as to why one would expect such factors of edge length in the observable.  Although the matter theory is massless, if one analyzes the classical field theory, one sees that the matter sector has a non-trivial energy-momentum tensor. In fact, the Hamiltonian for the system is:
\be
H =  \frac{1}{2}\frac{e^i_a e^j_b}{\sqrt{e^i e^i}}\epsilon_{ijk} \left (F[W]^k_{ab} +  \frac{i}{2}(\sigma^k)_{AB}\psi^A_a \psi^B_b\right) \quad\quad \textrm{where} \quad\quad e^i = \frac{1}{2}\epsilon_{ijk}\epsilon^{ab}e^j_a e^k_b.
\ee
We note at this point that the $e$-dependent prefactor has dimensions of length.  So from this argument, it comes as no surprise that the presence of matter should mean the insertion of a factor of length (that is, a factor linear in $k$) multiplying the holonomy.

Of course, there many possible combinations of insertion, but if we break them down into segments, then there are essentially three basic building blocks.  We are ready to draw these:

\begin{figure}[H]\centering
\begin{minipage}{0.4\linewidth}\centering
\psfrag{h}{}
\psfrag{g}{}
\psfrag{j}{}
\includegraphics[width =6cm]{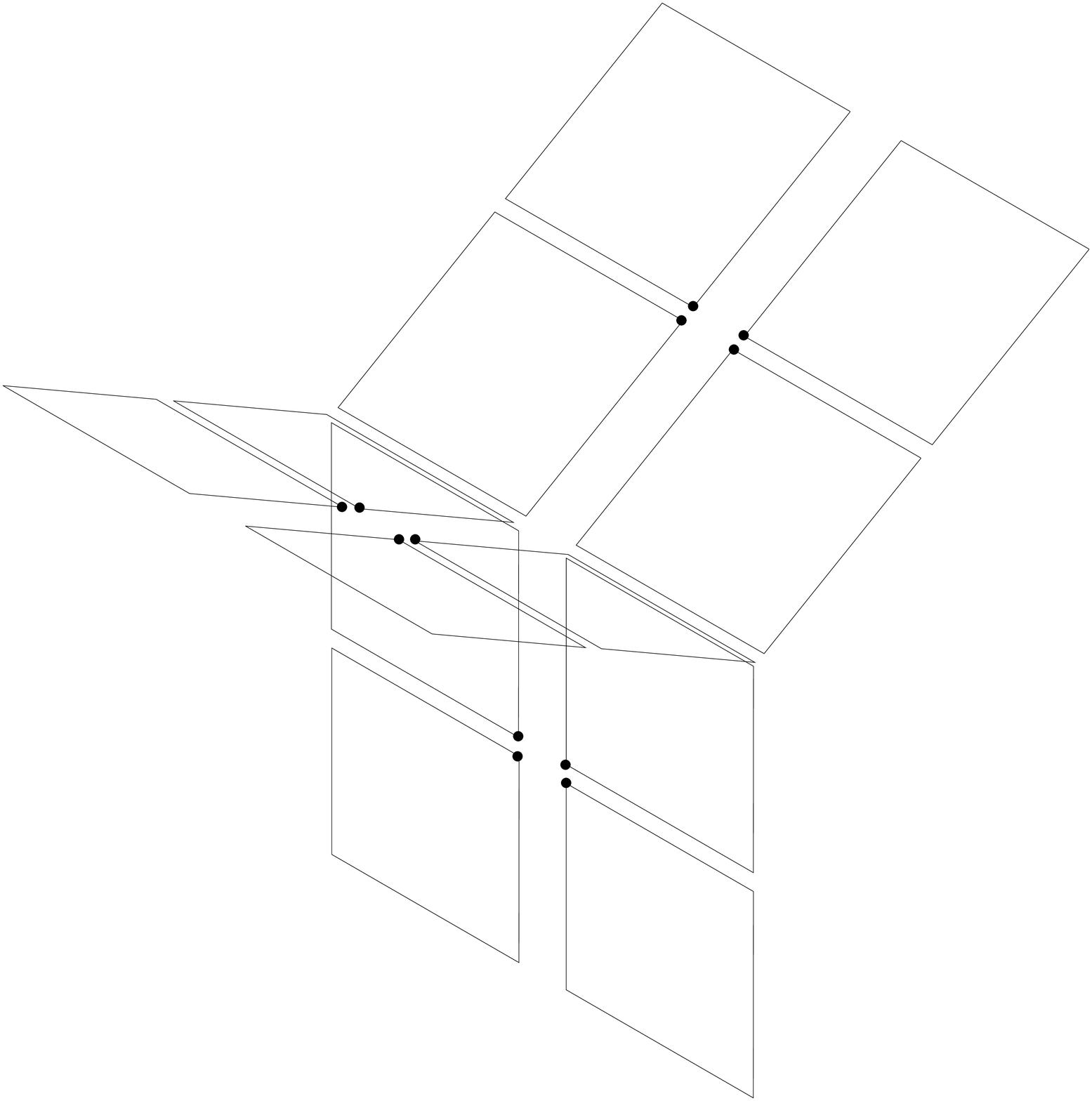}
\end{minipage}
\begin{minipage}{0.4\linewidth}\centering
\psfrag{h}{$g_{e^*_{e,f}}$}
\psfrag{g}{$g_{e^*_t}$}
\psfrag{j}{$j_{w^*}$}
\includegraphics[width =6cm]{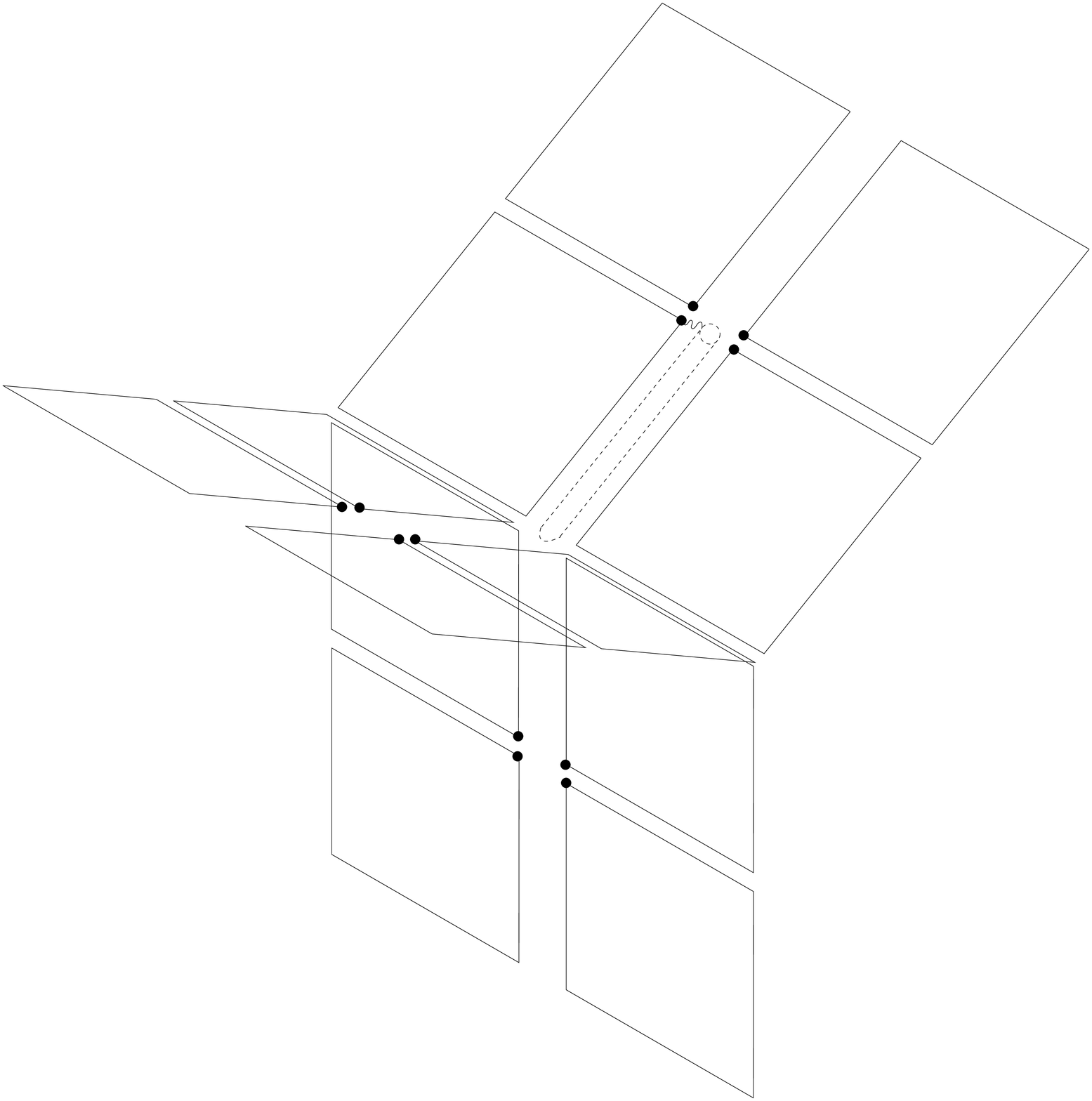}
\end{minipage}

\begin{minipage}{0.4\linewidth}\centering
\psfrag{h}{$g_{e^*_{e,f}}$}
\psfrag{g}{$g_{e^*_t}$}
\psfrag{j}{$j_{w^*}$}
\includegraphics[width =6cm]{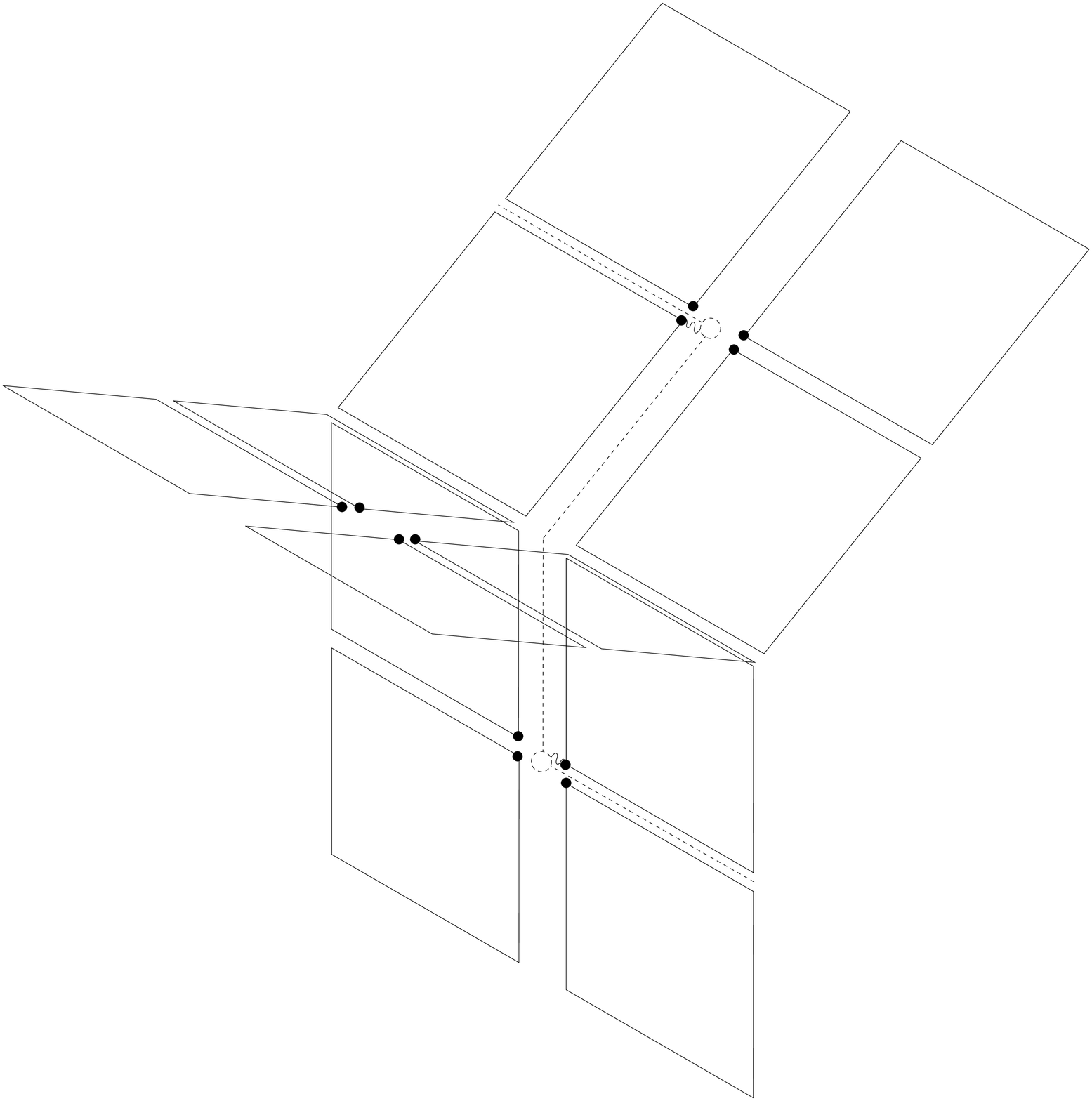}
\end{minipage}
\caption{\label{sampleone}Sample of the fermionic Feynman diagram insertions.}

\end{figure}

\noindent  We begin with the no-fermion case in the  top left. In going to the wedge formulation of the amplitude we actually already started to process of modifying the amplitudes.  The gravity projectors remove the factor of dimension  in $A_e$ since they project just onto the highest/lowest weight state and therefore kill the sum.  Furthermore, the factors of $(-1)^{2(j-k)(2j+1)}$ occurring in the projectors kill the same factors occurring in the triangle amplitudes $A_f$ . Notice that the gravity projectors are all at the center of the face.  This is to ensure that the final amplitude for the edges $e$ is correct.  The amplitudes for the various sub-simplices are now:
\be
\begin{array}{c}
\displaystyle A_{e} = (-1)^{2k_e},\qquad\qquad\qquad
A_{f} = (-1)^{ (k_{e_1} + k_{e_2} + k_{e_3})_f}\\
\\
\displaystyle A_{t} = \left[(-1)^{\sum_{a =1}^6 2(j_{e_a} - k_{e_a} )(2j_{e_a}+1)}
\left\{
\begin{array}{ccc}
k_{e_1} & k_{e_2} & k_{e_3}\vspace{0.1cm}\\
k_{e_4} & k_{e_5} & k_{e_6}
\end{array}
\right\}\right]_t.
\end{array}
\ee

Then there is the fermionic loop insertion, which also contributes to the bosonic sector of the theory, and which is illustrated in the top right.  The fermionic line traces a loop which is inserted between two wedges in one face.  Thus, the insertion does not map between $\SU(2)$ modules. In fact, using the standard retracing identity, we can remove the fermionic line altogether, but the non-trivial part is the clasp which  extracts a factor of $(-1)^{2(j-k)+1}(2k+1)$ for $V^k\subset R^j$.  Thus, the only difference between the no-fermion amplitude and the amplitude containing this fermion loop is the amplitude for one triangle:
\be
A_{f} = (-1)^{(k_{e_1}+ k_{e_2}+ k_{e_3})}(-1)^{2(j_{e_1} - k_{e_1}) +1}(2k_{e_1}+1),
\label{weight1}
\ee
where the loop was inserted around edge $e_1$.  
Let us take a specific example, say $k_{e_a} = j_{e_a}$ for all $a$, and let us sum up the four contributions: the no-insertion and the loop insertions on $e_1$, $e_2$, $e_3$.  The resulting amplitude is exactly $-A^{\{j\}}_f(\ua,\ua,\ua;\ua,\ua,\ua)$.
For the triangle given above, ultimately, the amplitude arising from summing over the no-insertion and the three possible  loop insertions will lead to all possible bosonic amplitudes depending on whether the edges are in the upper or lower modules. 

Let us move onto the fermionic contributions, which rely on non-trivial propagation of the fermion along edges of the simplicial complex. There is essentially one type of diagram, variations of which give the other 23 possibilities occurring in \eqref{amps1} and \eqref{amps2}.  We displayed the insertion in the center bottom of Fig.\ref{sampleone}.  We shall reproduce the following amplitude: $A^{\{j\}}_{f}(\ua,\ua,\ua;\ua,\da,\da)$.    This triangle amplitude is made up from several sub-diagrams:
\begin{figure}[H]\centering
\begin{minipage}{0.3\linewidth}\centering
\psfrag{a}{${}^{j_2}$}
\psfrag{b}{${}^{j_3}$}
\psfrag{c}{${}^{j_1}$}
\psfrag{d}{${}^{j_2 - \frac{1}{2}}$}
\psfrag{e}{${}^{j_3 - \frac{1}{2}}$}
\psfrag{r}{${}^{-1}$}
\psfrag{s}{${}^{\frac{1}{2}}$}
\includegraphics[height = 5cm]{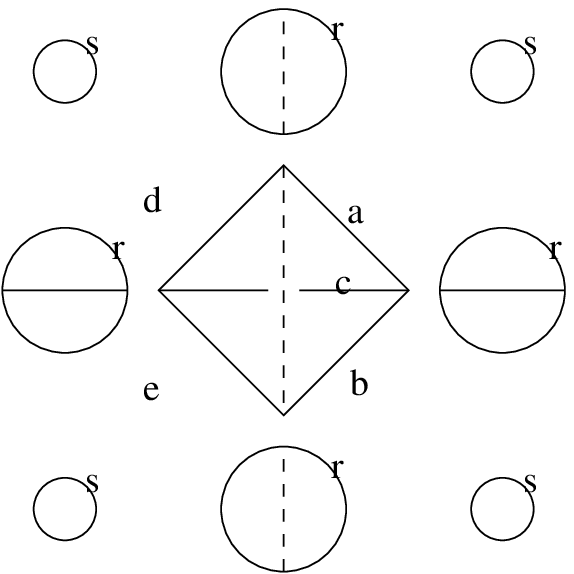}
\end{minipage}
\begin{minipage}{0.6\linewidth}
\be\quad= \quad
(-1)^{j_1+j_2+j_3  +(2j_1 + 1) }\sqrt{(j_1+j_2+j_3+1)(-j_1+j_2+j_3)}
\label{weight2}
\ee
\end{minipage}
\end{figure}
\noindent which we note is just minus the amplitudes for which we were hoping.  We stress that the diagrams in this equation all contribute to the triangle amplitude $A_f$.  We get a tetrahedral diagram, because we have coupled an extra spin-$\frac{1}{2}$ between the wedges.  The square root of the loops come from factors of  dimension which we saw, in the gravity case, arise upon decomposition of the $\delta$-functions.  Therefore, we have successfully reproduced the triangle amplitudes, which is what was our goal.

Thankfully, this allows for a thorough description of all the supersymmetric amplitudes.  All we need is to take the pure gravity ampliudes and insert all the possible fermionic observables consistent with the above rules and as such we arrive at the supersymmetric amplitude.

\begin{figure}[H]\centering
\begin{minipage}{0.4\linewidth}\centering
\includegraphics[height = 4cm]{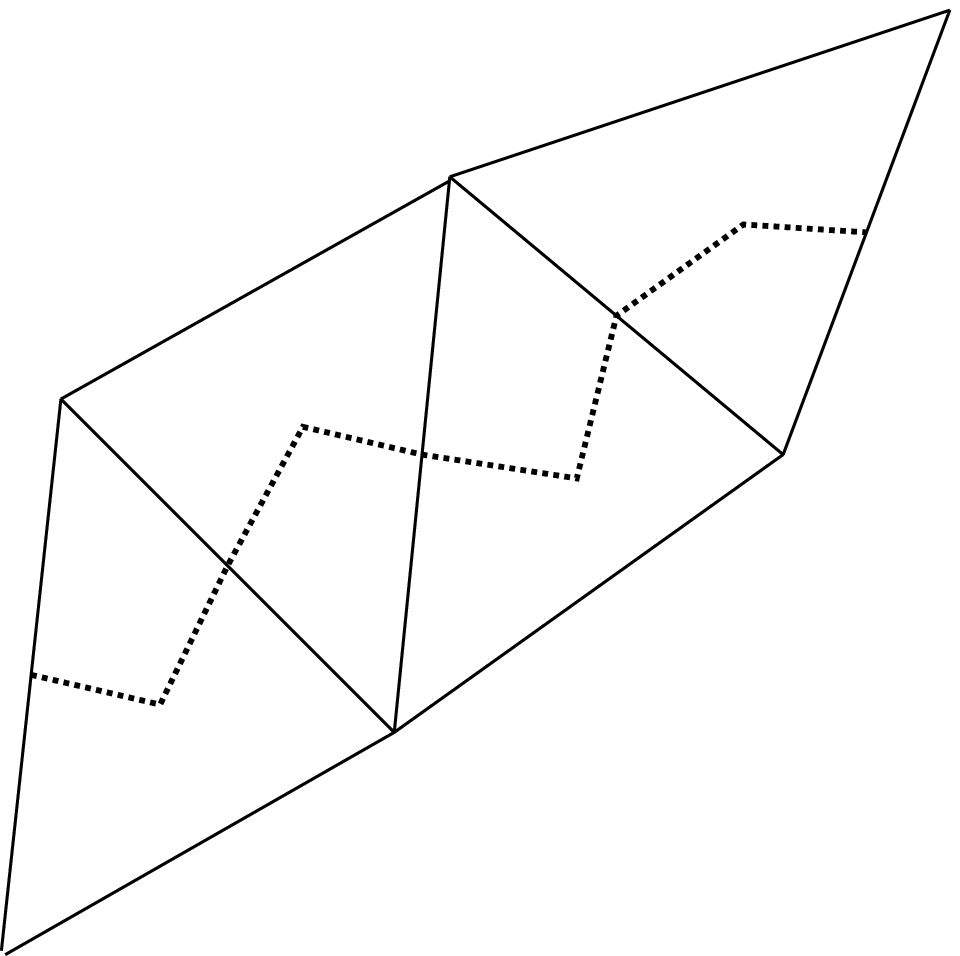}
\end{minipage}
\begin{minipage}{0.1\linewidth}
$\longrightarrow$
\end{minipage}
\begin{minipage}{0.4\linewidth}
\includegraphics[height = 4cm]{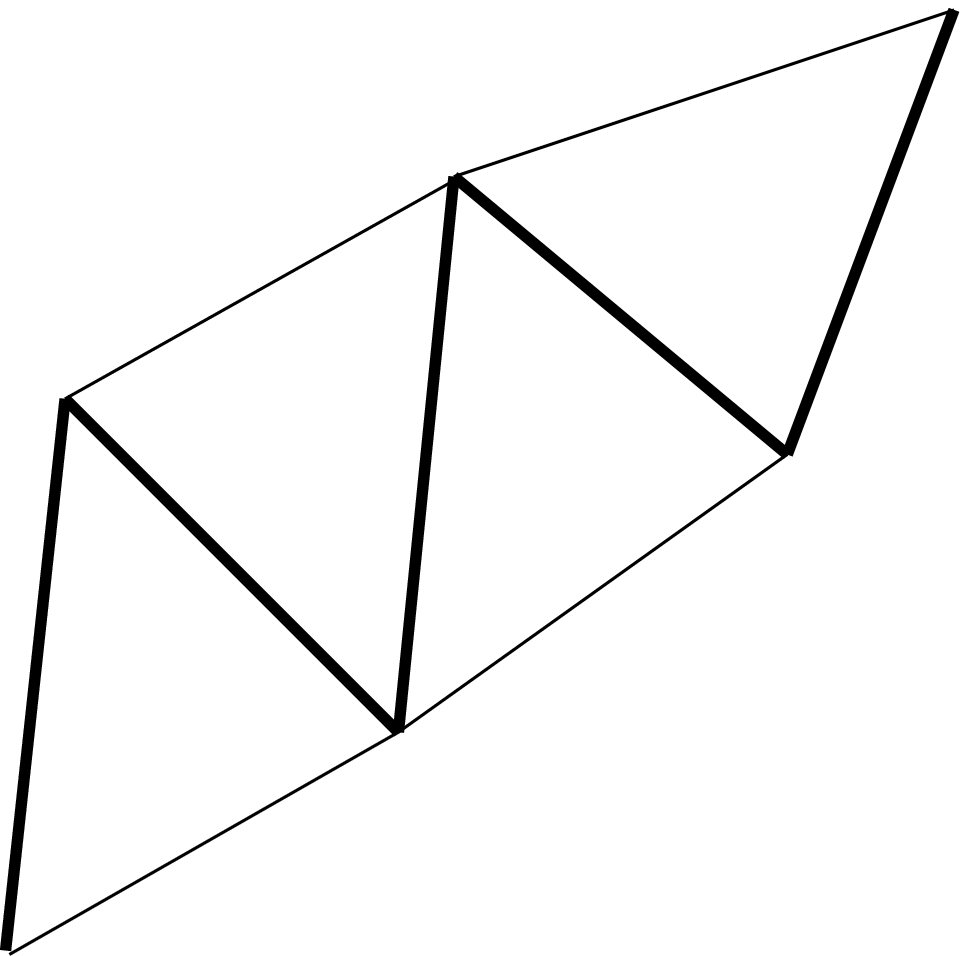}
\end{minipage}
\caption{\label{feynmand}From spin foam to spacetime picture.}
\end{figure}

As promised, we conclude with a description of the geometric properties of the fermionic observables.  consider two adjacent wedges.  Their intersection is an edge $e^*_{e,f}$, joining the center of a triangle $f\subset\Delta$ with the midpoint of an edge $e\subset\Delta$.  The path of the fermion in the spin foam, the dashed line, contains only such edges (see Fig.\ref{feynmand} for details).  The obvious spacetime picture is that the particle propagates along the edges $e$ occurring in $e^*_{e,f}$.  This is a perfectly self-consistent propagation and gives a nice geometric viewpoint to the amplitude.

At the end of the day, putting aside the trivial fermionic loop insertion of the top-right type in Fig.\ref{feynmand}, a fermionic Feynman diagram is a graph embedded in the triangulation. More precisely, we want sets of  closed loops of edges, such that two consecutive edges on a loop belong to a same triangle. This means that we can equivalently think of the fermionic loop as a closed sequence of triangles i.e. a loop in the spinfoam complex. Then considering a given edge $e$ propagating a fermion, and looking at the plaquette around it, the triangle that contains two fermionic edges will trigger a $\f12$-shift of $k_e$. This will happens twice around the plaquette, once by a $+\f12$-shift, once by a $-\f12$-shift. These two triangles correspond to the two triangles of the Feynman diagram sharing the same edge. 
Now, we can have an arbitrary number of fermionic loops in the Feynman diagram and actually  they can share the same edges, since there is no fermionic interaction term here, so that the only interactions are between the fermionic field(s) and the gravitational degrees of freedom.  Thus, in general there can be several pairs of $\pm\f12$-shifts as one goes around a plaquette, each corresponding to a separate Feynman diagram.
The last step is to sum over all possible Feynman diagrams in order to reconstitute the full supersymmetric spinfoam amplitude.

This picture is finally slightly different from the one initially envisioned in \cite{livineoeckl}.

\section{Conclusions}


Starting from the topological spinfoam model for $N=1$ supergravity in 3d gravity, we have analyzed in detail the structure of these spinfoam amplitudes. We have first shown how to derive these spinfoam amplitudes from a discretised BF action on a triangulation by extending the standard bosonic construction of a discretised action for the $\SU(2)$ Ponzano-Regge model \cite{PR1}  to include for fermionic degrees of freedom in the connection and triad. In particular, this showed how including fermions can resolve the standard ambiguity that the usual discretised action leads to $\SU(2)/\Z_2\sim\SO(3)$ and not exactly to $\SU(2)$.

Then we explicitly  decomposed all supersymmetric amplitudes into a superposition of the standard $\SU(2)$ amplitudes. This is done by decomposing $\UOSP(1|2)$ representations into irreducible representations of $\SU(2)$. The most striking result is that although a single spin $j_e$ is associated to each edge of the triangulation, the actual length of that edge is a priori different seen from the viewpoint  of each tetrahedron to which it belongs: it can be either $k_e=j_e$ or $k_e=j_e-\f12$ depending whether a fermion is traveling through this tetrahedron or not. Pushing this decomposition into $\SU(2)$ amplitudes as far as  possible, we finally showed that the supersymmetric amplitude can be seen as the coupling of fermionic Feynman diagrams to the gravitational background. Let us emphasise that the geometry  is not static but when a fermionic line is inserted, it creates length shifts as mentioned previously.

If we were to go further in the understanding of these $N=1$ supersymmetric spinfoam models, we could analyze the asymptotics of the susy $\{6j\}$-symbol and see how the Regge action and the fermionic fields appears in the large spin limit \cite{inprep}. We should also compare our approach to the standard insertion of particles with spin in the Ponzano-Regge model \cite{PR1} (the actual difference is that our framework takes into account explicitly the feedback of the fermionic fields on the gravitational fields) and to the more recent gravity+fermions models developed in \cite{winston1,winston2}.

Finally, the most interesting application to our formalism would be to study the insertion of actual physical non-topological fermionic fields.  Starting in 3d, in the present work, we have tracked from the initial continuum action down to the final discretised spinfoam amplitude how the explicit fermionic Feynman diagrams get inserted in the spinfoam amplitude. These fermionic observables come with precise weights (see e.g. eqn. \eqref{weight1}-\eqref{weight2}). These weights are fine-tuned so as to ensure that the full model `gravity+fermions' is topological. That shows that these spinfoam amplitudes provide the correct quantisation for our supersymmetric theory. As soon as we modify these weights, we would get non-topological amplitudes and it would be interesting to see how we could modify them in order to insert more physical fermionic fields. Then, we hope to apply the same procedure to the four-dimensional case by first deriving the spinfoam quantisation of supersymmetric BF theory and studying how the fermions are coupled to the spinfoam background, and then seeing how this structure is maintained or deformed when we introduce the (simplicity) constraints on the $B$-field  in order to go from the topological BF theory down back to proper gravity. Another interesting outlook is to push our analysis to $N=2$ supersymmetric BF theory, already in three space-time dimensions, following the footsteps of \cite{Livine:2007dx}. Indeed, such a theory already include a spin-1 gauge field, and we could study in more detail how the full supersymmetric amplitudes decomposes into Feynman diagrams for the fermions and spin-1 field inserted in the gravitational spinfoam structure. Then we would see how it is possible to deform this structure in such a way that the spin-1 field represents standard gauge fields. This road would provide an alternative way to coupling (Yang-Mills) gauge fields to spinfoam models, which we could then compare to the other approaches developed in this direction \cite{YM}.

\section*{Acknowledgements}
VB and JR would like to acknowledge the support of the researchers and staff of Perimeter Institute, where a large part of this work was completed.

\begin{appendix}

\section{Mathematical preliminaries}

\subsection{Grassmann algebra}
\label{grass}

A Grassmann algebra $\cG$,  with $N$ generators $\xi_1, \dots,\, \xi_N$ satisfies $\xi_i\xi_j + \xi_j\xi_i = 0$.  ($N$ may be both finite and infinite.)  The element:
\be
\label{grassy}
\alpha = \sum_{m\geq 0} \sum _{i_1<\dots<i_m} \alpha_{i_1\dots i_m}\xi_{i_1} \xi_{i_m},
\ee
is called {\it even} if only the coefficients with even $m$ are non-zero and {\it odd} if only the coefficients with odd $m$ are non-zero.  The sets of even and odd elements are denoted $\cG_0$ and $\cG_1$ respectively, and $\cG = \cG_0\oplus\cG_1$.

The parity function $\lambda(\alpha)$ is defined on $\cG$ as:
\be
\lambda(\alpha) = \left\{
\begin{array}{ccl}
0 &\text{if} & \alpha\in\cG_0,\vspace{0.1cm}\\
1 &\text{if} & \alpha\in\cG_1.
\end{array}\right.
\ee
We define a {\it complex conjugation} operation, $\square$, on $\cG$ with the following properties:
\be
(\alpha\beta)^{\square} = \alpha^\square\beta^\square, \qquad (c\, \alpha)^\square = \bar{c}\, \alpha^\square, \qquad (\alpha^\square)^\square = (-1)^{\lambda(\alpha)}\alpha,
\ee
where $\alpha,\,\beta\in\cG$ and $c\in\mathbb{C}$ and $\bar{c}$ denotes standard complex conjugation on $\mathbb{C}$.  There are a number of ways to define such an operation on a Grassmann algebra, so one must pick one and adhere to it.

We can define an integration theory for functions $f: \cG\rightarrow \cG$.  First of all, analytic functions on $\mathbb{C}$ have a natural extension to superanalytic functions on $\cG$:
\be
f(\alpha) = \sum_{n\geq 0} \frac{f^{(n)}(\alpha_*)}{n!}(\alpha -\alpha_*)^n,
\ee
where $\alpha_*$ is known as the {\it body} of $\alpha$ in DeWitt's terminology \cite{dewitt}. It is the $m = 0$ term in (\ref{grassy}) (while the remainder $\alpha - \alpha_*$ is its {\it soul}).  We can define a measure to integrate functions on $\cG$.  For {\it even} elements, the {\it body} plays a special role:
\be
\alpha = \alpha_* + \sum_{m\geq 2} \sum _{i_1<\dots<i_m} \alpha_{i_1\dots i_m}\xi_{i_1}\dots \xi_{i_m},
\ee
 Then, the measure is:
\be
\int d\alpha := \int d\alpha_*,
\ee
where one also replaces $\alpha$ by $\alpha_*$ in the integrand.  Clearly, {\it odd} elements have no {\it body}.  Thus, we need a different definition of the measure.  The most general superanalytic function of an {\it odd} element is: $f(\alpha) = (c_1 +c_2\alpha)$.  For {\it odd} elements, the measure is defined as:
\be
\label{grassyodd}
\int d\alpha\; (c_1 +c_2\alpha) = c_2, \quad\quad\text{which means that} \quad\quad \int d\alpha \;\alpha\,f(\alpha) = f(0),
\ee
where $\delta(\alpha_*)$ is the standard distributional one on $\mathbb{C}$.  Hence, we can define a delta function on $\cG$.  For the functions on $\cG_0$:
\be
\int d\alpha\; f(\alpha)\, \delta(\alpha) := \int d\alpha_*\; f(\alpha_*)\, \delta(\alpha_*) = f(0),
\ee
and for functions on $\cG_1$, we can see that $\delta(\alpha) := \alpha$, as can be seen in (\ref{grassyodd}).

\subsection{Super Lie algebra $\osp(1|2)$}
\label{osp}

The algebra $\osp(1|2)$ is a super Lie algebra \cite{Scheunert:1976wj, bertol}.   There is a parity function defined on $\osp(1|2)$ which
divides its elements into {\it even} and {\it odd} subsets:
\be
\lambda(X) = \left\{
\begin{array}{ccl}
0 & \text{if} & X\in\osp(1|2)_0\vspace{0.1cm}\\
1 & \text{if} & X\in\osp(1|2)_1.
\end{array}\right.
\ee
The set $\osp(1|2)_0 \sim \su(2)$ contains three generators $J_1,\, J_2,\, J_3$, while the set $\osp(1|2)_1$ contains two generators $Q_\pm$.

We define a bracket on this algebra by:
\be
[X_1,\, X_2] = (-1)^{\lambda(X_1)\lambda(X_2)+1} [X_2, X_1]
\ee
and which satisfies a super Jacobi identity.\footnote{The super Jacobi identity is:
\be
(-1)^{\lambda(X_1)\lambda(X_3)}[X_1,[X_2,X_3]] + (-1)^{\lambda(X_2)\lambda(X_1)}[X_2,[X_3,X_1]] + (-1)^{\lambda(X_3)\lambda(X_2)}[X_3,[X_1,X_2]]  = 0.
\ee}  All together, the generators satisfy the algebra:
\be{
\begin{array}{lcll}
[J_3,J_\pm] = \pm i J_\pm, &\quad  & [J_+,J_-] = 2iJ_3,& \vspace{0.3cm}\\

[J_3,Q_\pm]= \pm\frac{i}{2} Q_\pm, &\quad &   [J_\pm,Q_\pm] = 0,  &  [J_\pm, Q_\mp] = iQ_\pm,\vspace{0.3cm}\\

 [Q_\pm,Q_\pm] = \mp \frac{i}{2} J_\pm,&\quad &  [Q_\pm, Q_\mp] = \frac{i}{2} J_3&
\end{array}}
\ee

We define a {\it supertranspose} operation $\ddagger$ on $\osp(1|2; \mathbb{C})$, which has the properties:
\be
(c_1X_1 + c_2 X_2)^\ddagger = \bar{c}_1X_1^\ddagger + \bar{c}_2X_2^\ddagger, \qquad [X_1,X_2]^\ddagger = (-1)^{\lambda(X_1)\lambda(X_2)}[X_2^{\ddagger}, X_1^{\ddagger}],  \qquad  (X^\ddagger)^\ddagger = (-1)^{\lambda(X)}X^\ddagger,
\ee
where $X_i \in \osp(1|2)$ and $c_i \in \mathbb{C}$.   There are two such operations for the generators of $\osp(1|2)$:
\be
J_i^{\ddagger} = -J_i, \qquad
Q_+^\ddagger = (-1)^{\epsilon}\, Q_-, \qquad
Q_-^\ddagger = (-1)^{\epsilon+1}\, Q_+,
\ee
for $\epsilon = 0,\, 1$.

We define a {\it grade adjoint} operation $\dagger$ on $\osp(1|2; \cG) = \osp(1|2; \cG_0)_0\oplus\osp(1|2; \cG_1)_1$:
\be
(\alpha_1X_1 + \alpha_2X_2)^\dagger = \alpha_1^\square X_1^\ddagger + \alpha_2^\square X_2^\ddagger.
\ee
For our purposes, we confine to a subalgebra $\tilde{\cG}\subset \cG$, such that every element of $\uosp(1|2) := \osp(1|2; \tilde{\cG})$ satisfies $X^\dagger = -X$.  Therefore, depending on the choice of grade adjoint operation, the elements are of the form:
\be
X = \alpha_1\, J_1 + \alpha_2 \,J_2 + \alpha_3\, J_3 + (-1)^\epsilon\alpha^\square\, Q_+ + \alpha\,Q_-,
\ee
where $\alpha_i^\square = \alpha_i$ and $(\alpha^\square)^\square = -\alpha$.

\vspace{1cm}

Primarily, the representations of $\uosp(1|2)$ are labelled by a half-integer $j$ and a parity $\lambda \in \{ 0,\,1\}$. One has a certain freedom as to the inner product one chooses for a given representation, which is parametrised by two more numbers $\rho, \tau \in \{0,\, 1\}$.  We shall denote such a representation by: $R^{j,\lambda,\rho,\tau}$.  These representations can be decomposed over the even subalgebra: $\uosp(1|2)_0 \simeq \su(2)$.  Each representation $R^{j,\lambda,\rho,\tau}$ of $\uosp(1|2)$ comprises of the direct sum of two representations of $\su(2)$.
\be
R^{j,\lambda,\rho,\tau} = V^{j,\lambda,\rho,\tau} \oplus V^{j-\frac{1}{2},\lambda+1,\rho, \tau}.
\ee
A generic basis element is $|j;k,m> $, where $k \in \{ j,\, j-\frac{1}{2}\}$,  $m\in\{-k,\,-k+1,\,\dots,\, k\}$ and we have suppressed the labels $\lambda, \rho, \tau$ for simplicity. The action of the operators on  the representation $R^{j,\lambda,\rho,\tau}$ is:
\be
\begin{array}{rcl}
J_3|j;k,m> &=& i m\,|j;k,m> , \vspace{0.4cm}\\

J_\pm |j;j,m> &=& i\sqrt{(j\mp m)(j\pm m + 1)} |j;j, m\pm 1>, \vspace{0.4cm}\\
 J_\pm |j;j-\frac{1}{2},m> &=& i\sqrt{(j-\frac{1}{2}\mp m)(j+\frac{1}{2}\pm m)} |j;j-\frac{1}{2}, m\pm 1>,\vspace{0.4cm}  \\

Q_{\pm} |j;j,m>  &=&  \mp\frac{1}{2}\sqrt{j\mp m}|j;j-\frac{1}{2}, m\pm \frac{1}{2}>, \vspace{0.4cm} \\
 Q_{\pm} |j; j-\frac{1}{2}, m>  &=&  -\frac{1}{2}\sqrt{j+\frac{1}{2} \pm m}\;|j;j,m\pm \frac{1}{2}>,
\end{array}
\ee
where $J_\pm := J_i \pm iJ_2$ and $|j;k,m>$ has parity $\lambda + 2(j-k)$.  The inner product, $\Phi^{(\rho,\tau)}(\;\cdot\;,\;\cdot\;)$, on such a representation is defined by:
\be
\Phi^{(\rho,\tau)}\left(|j;k,m>, |j;k',m'>\right) := (-1)^{2(j-k)\rho+\tau}  \delta^{kk'}\delta^{m}{}_{m'} = (-1)^{\varphi}\delta^{kk'}\delta^{m}{}_{m'},
\ee
where we define $\varphi := 2(j-k)\rho +\tau$ for later convenience.
One can show that there is a consistency relation among $\epsilon$, $\lambda$ and $\rho$:
\be
\epsilon + \lambda + \rho + 1 \equiv 0 \quad \textrm{(mod 2)}.
\ee
Therefore, once two of these parameters are chosen, the final one is fixed.  From a spin-statistics viewpoint, we would like to endow the integer representations with even parity and the half-odd-integer representations with odd parity, that is, $\lambda \equiv 2j \; \textrm{(mod 2)}$.   Furthermore, in the near future, we shall wish to define a measure on the supergroup that can be used to integrate all representation functions.  Thus, we must have just one definition of grade adjoint; we shall choose $\epsilon = 0$.  We conclude that $\rho \equiv 2j+1 \; \textrm{(mod 2)}$. Ultimately, we are free with our choice of the overall sign $\tau$, {\it but} we must include both choices.  The reason for this will appear shortly when we consider tensor products of representations.  In fact, these choices mean that for integer representations, one does not acquire a positive definite inner product on the representation space:
\be<j,\tau;k,m|j,\tau;k',m'> = (-1)^{\varphi}  \delta^{kk'}\delta^{m}{}_{m'} =
\left\{
\begin{array}{rcl}
(-1)^{2(j-k)+ \tau}\delta^{kk'}\delta^{m}{}_{m'} &\quad\textrm{for}& \quad j\in \mathbb{N},\\
 (-1)^\tau \delta^{kk'}\delta^{m}{}_{m'} &\quad\textrm{for}& \quad j\in\mathbb{N}+\frac{1}{2}.
\end{array}\right.
\ee
Furthermore, the tensor product of two representations of $\uosp(1|2)$ satisfies a rule analogous to that of $\su(2)$ except that the sum over $j$ goes in {\it half-integer} steps
\be
R^{j_1, \tau_1}\otimes R^{j_2,\tau_2} = \bigoplus_{|j_1 - j_2| \leq j_3 \leq j_1 + j_2} R^{j_3(j_1,j_2),\tau_3(\tau_1,\tau_2)}.
\ee
Using the properties of the inner product on the representation space, we find that:
\be
\tau_3(\tau_1,\tau_2) = \tau_1 + \tau_2 + 2(j_1+j_2+j_3)\lambda_3 + \lambda_1\lambda_2.
\ee
Thus, we see our initial requirement that we include both values of $\tau$ is justified; we cannot restrict to one particular choice of $\tau$ since we will obtain both under tensor composition.
We also choose that in the matrix realisation:
\be
{}^{j,\tau}T^{(km)}{}_{(ln)}(X) :=\; <j,\tau;k,m|X|j,\tau;l,n>\; =
\left(
\begin{array}{r|l}
\uosp(1|2)_0\leftarrow\uosp(1|2)_0 &  \uosp(1|2)_0\leftarrow\uosp(1|2)_1\\
\hline
\uosp(1|2)_0\leftarrow\uosp(1|2)_1 & \uosp(1|2)_1\leftarrow\uosp(1|2)_1
\end{array}
\right).
\ee
The {\it supertrace} of a matrix operator ${}^{j,\tau}\!M^{(km)}{}_{(ln)}$ (in the representation $R^{j,\tau}$) is defined as:
\be
\Str({}^{j,\tau}\!M) = \sum_{k,m} (-1)^{\lambda + 2(j - k)}\;\;{}^{j,\tau}\!M^{(km)}{}_{(km)} = \sum_{k,m} (-1)^{2k}\;\;{}^{j,\tau}\!M^{(km)}{}_{(km)}.
\ee
This means that the {\it supertrace} of the identity operator is: $\Str({}^{j,\tau}\id) = (-1)^{2j}$.  With these choices, the matrix elements of the generators in the fundamental representation  $R^{\frac{1}{2},0} = V^{0} \oplus V^{\frac{1}{2}}$ are:
\be
\begin{split}
J_1 = \frac{i}{2}\left(
\begin{array}{ccc}
0 & 0 & 0\\
0 & 0 & 1\\
0 & 1 & 0\\
\end{array}\right), \quad
J_2 = \frac{i}{2}\left(
\begin{array}{ccc}
0 & 0 & 0\\
0 & 0 & -i\\
0 & i & 0\\
\end{array}\right), \quad
J_3 = \frac{i}{2}\left(
\begin{array}{ccc}
0 & 0 & 0\\
0 & 1 & 0\\
0 & 0 & -1\\
\end{array}\right), \\
Q_+ = \frac{1}{2}\left(
\begin{array}{ccc}
0 & 0 & -1\\
-1 & 0 &  0\\
0 & 0 & 0\\
\end{array}\right), \quad
Q_- = \frac{1}{2}\left(
\begin{array}{ccc}
0 & 1 & 0\\
0 & 0 & 0\\
-1 & 0 & 0\\
\end{array}\right).\hphantom{xxxxxxxxx}
\end{split}
\ee
The {\it supertrace} in the fundamental representation is:\footnotemark
\footnotetext{The spinor indices follow the north-west convention so that $\phi^{A} = \epsilon^{AB}\phi_B$ and $\phi_{A} = \phi^B\epsilon_{BA}$.  The metric on the spinor space is the anti-symmetric tensor  $\epsilon_{AB}$ with $\epsilon_{+-} = \epsilon^{+-} = 1$.  This implies $\epsilon^{AB}\epsilon_{BC} = - \delta^A{}_{C}$.}
\be
\Str(J_iJ_j)  = \frac{1}{2}\delta_{ij}, \quad\quad \Str(J_iQ_A) = 0, \quad\quad \Str(Q_AQ_B) = \frac{1}{2} \epsilon_{AB}.
\ee
The measure over the algebra is:
\be
dB = db_1\, db_2\, db_3\, db^\square\, db,
\ee
where $B = b_iJ_i + b^\square Q_+ + bQ_-$.

\subsection{Super group $\UOSP(1|2)$}
\label{OSP}

Elements of $\UOSP(1|2)$ have the form:
\be
g = u\xi, \qquad\qquad \text{where} \qquad\qquad u = e^{\theta \vec{n}\cdot\vec{J}} \quad \text{and} \quad \xi = e^{\eta^\square Q_+ + \eta Q_-},
\ee
and $\vec{n} = (\sin\psi\cos\phi,\, \sin\psi\sin\phi,\, \cos\psi)$.  More explicitly:
\be
u= \left(
\begin{array}{ccc}
	1 & 0 & 0\vspace{0.2cm}\\
	0 &\cos\theta + \frac{i}{2}\sin\theta\cos\psi &   \frac{i}{2}\sin\theta\sin\psi\, e^{-i\phi}\vspace{0.2cm} \\
	 0 & \frac{i}{2}\sin\theta\sin\psi\, e^{i\phi} & \cos\theta - \frac{i}{2}\sin\theta\cos\psi
\end{array}\right)\qquad \text{and} \qquad
\xi= \left(
\begin{array}{ccc}
	1+\frac{1}{4}\eta^\square\eta & \frac{1}{2}\eta & -\frac{1}{2}\eta^\square\vspace{0.2cm}\\
	-\frac{1}{2}\eta^\square &1-\frac{1}{8}\eta^\square\eta & 0  \vspace{0.2cm}\\
	 -\frac{1}{2}\eta & 0 & 1-\frac{1}{8}\eta^\square\eta   			
\end{array}\right)
\ee
Thus, we arrive at:
\be
g= \left(
\begin{array}{ccc}
	1+\frac{1}{4}\eta^\square\eta & \frac{1}{2}\eta & -\frac{1}{2}\eta^\square\vspace{0.2cm}\\
	g_{21} & (1-\frac{1}{8}\eta^{\Box}\eta)u_{22} &  (1-\frac{1}{8}\eta^{\Box}\eta)u_{23} \vspace{0.2cm}\\
	g_{31} &(1-\frac{1}{8}\eta^{\Box}\eta)u_{32} & (1-\frac{1}{8}\eta^{\Box}\eta)u_{33}
\end{array}\right)
\qquad \text{with} \qquad\quad
\begin{array}{rcl}
 g_{21} &=& -\frac{1}{2}\eta^\square u_{22}-\frac{1}{2}\eta u_{23},\vspace{0.2cm}\\
 g_{31} &=& -\frac{1}{2}\eta^\square u_{32}-\frac{1}{2}\eta u_{33},
 \end{array}
\ee
Interestingly, elements of group $\UOSP(1|2)$ satisfy the relations $g^\dagger g = gg^\dagger= \id$ and $g^{\ddagger} \zeta g = \zeta$, where $\zeta := diag(\id_{1x1},\epsilon_{2x2} )$.\footnote{The {\it grade adjoint} and {\it supertranspose} are:
\be
g^\dagger= \left(
\begin{array}{ccc}
	1+\frac{1}{4}\eta^\square\eta & g_{31} & -g_{21}\vspace{0.2cm}\\
	\frac{1}{2}\eta^\square & (1-\frac{1}{8}\eta^{\Box}\eta)u_{33} &  -(1-\frac{1}{8}\eta^{\Box}\eta)u_{23} \vspace{0.2cm}\\
	\frac{1}{2}\eta &-(1-\frac{1}{8}\eta^{\Box}\eta)u_{32} & (1-\frac{1}{8}\eta^{\Box}\eta)u_{33}
\end{array}\right)
\quad\quad
\textrm{and}\quad\quad
g^\ddagger= \left(
\begin{array}{ccc}
	1+\frac{1}{4}\eta^\square\eta & -g_{21} & -g_{31}\vspace{0.2cm}\\
	\frac{1}{2}\eta & (1-\frac{1}{8}\eta^{\Box}\eta)u_{22} &  (1-\frac{1}{8}\eta^{\Box}\eta)u_{32}\vspace{0.2cm} \\
	-\frac{1}{2}\eta^\square &(1-\frac{1}{8}\eta^{\Box}\eta)u_{23} & (1-\frac{1}{8}\eta^{\Box}\eta)u_{22}
\end{array}\right).
\ee
}
From the second relation, we can see the origin of the description orthosymplectic.

\vspace{0.5cm}

The representation matrices of the group elements are denoted ${}^{j,\tau}T^{(k\,m)}{}_{(l\,n)}(g) = {}^{j,\tau}T^{(k\,m)}{}_{(l\,n)}(\Omega, \eta^\square, \eta) $, where $\Omega = \{\psi,\, \theta,\, \phi\}$ and have elements:
\be
\label{rep}
\begin{array}{rcl}
{}^{j,\tau}T^{(j\,m)}{}_{(j\, n)} (g)
				&=&(-1)^{\tau}(1-\frac{1}{4}j\,\eta^\square\eta)\; {}^j\!D^{m}{}_{n}(\Omega), \vspace{0.3cm}\\
{}^{j,\tau}T^{(j\, m)}{}_{(j-\frac{1}{2}\, n)} (g)
				&=&(-1)^\tau\left[-\frac{1}{2}\sqrt{j+n+\frac{1}{2}}\;\eta^\square\; {}^j\!D^{m}{}_{n+\frac{1}{2}}(\Omega) - \frac{1}{2}\sqrt{j-n+\frac{1}{2}}\;\eta\; {}^j\!D^{m}{}_{n-\frac{1}{2}}(\Omega)\right],\vspace{0.3cm}\\
{}^{j,\tau}T^{(j-\frac{1}{2}\, m)}{}_{(j\, n)} (g)
				&=&(-1)^{\rho + \tau}\left[-\frac{1}{2}\sqrt{j-n}\; \eta^\square\;\; {}^{(j-\frac{1}{2})}\!D^{m}{}_{n+\frac{1}{2}}(\Omega) + \frac{1}{2}\sqrt{j+n}\;\eta\;\; {}^{(j-\frac{1}{2})}\!D^{m}{}_{n-\frac{1}{2}}(\Omega)\right], \vspace{0.3cm}\\

{}^{j,\tau}T^{(j-\frac{1}{2}\,  m)}{}_{(j-\frac{1}{2}\, n)} (g)
				&=&(-1)^{\rho+\tau}(1+\frac{1}{4}(j+\frac{1}{2})\,\eta^\square\eta)\; {}^{(j-\frac{1}{2})}\!D^{m}{}_{n}(\Omega),
\end{array}
\ee
where $\rho = 2j+ 1$ and $D^j(\Omega)$ is the $j$th representation of the $\SU(2)$ group element pertaining to $\Omega$.  Matrix multiplication, {\it complex conjugation}  and the {\it grade adjoint} operation satisfy the following relations:
\be
\begin{array}{rclcl}
{}^{j,\tau}T^{(km)}{}_{(ln)}(g) & = &  {}^{j,\tau}T^{(km)}{}_{(ln)}(g_1g_2) &= &  {}^{j,\tau}T^{(km)}{}_{(k'm')}(g_1)\;{}^{j,\tau}T^{(k'm')}{}_{(ln)}(g_2)\vspace{0.3cm}\\
{}^{j,\tau}T^{(km)}{}_{(ln)}(g)^\square  &=& {}^{j,\tau}T^{(km)}{}_{(ln)}(g^\square) &= &{}^{j,\tau}T_{(km)}{}^{(ln)}(g) \vspace{0.3cm}\\
{}^{j,\tau}T^{(km)}{}_{(ln)}(g)^\dag  &=& {}^{j,\tau}T^{(km)}{}_{(ln)}(g^\dag) &= & {}^{j,\tau}T^{(l n)}{}_{(k m)}(g)
\end{array}
\ee
Indices are raised and lowered using a metric and its inverse:
\be\label{OSP6}
\begin{array}{rcl}
{}^{j,\tau}r^{(km)(ln)} & = & (-1)^{2(j - k)(2j+1) + k - m}\; \delta^{kl} \delta^{m+n, 0},\\
{}^{j,\tau}r_{(km)(ln)} & = & (-1)^{2(j - l)(2j+1) + l - n} \;\delta^{kl} \delta^{m+n, 0}.
\end{array}
\ee
The representation functions are orthogonal:
\be
\int_{\UOSP(1|2)} dg \;{}^{j_1,\tau_1}T^{(k_1m_1)}{}_{(l_1n_1)}(g)\;{}^{j_2,\tau_2}T_{(k_2m_2)}{}^{(l_2n_2)}(g) = \delta^{j_1j_2}\delta^{\tau_1\tau_2}\;\;  {}^{j_1,\tau_1}\delta^{(k_1m_1)}{}_{(k_2m_2)}\; {}^{j_1,\tau_1} \delta_{(l_1n_1)}{}^{(l_2n_2)}
\ee

\section{Super $\{3j\}$-symbols}
\label{intertwiners}

We can derive the Clebsch-Gordan coefficients quite easily from relations given above \cite{Daumens:1992kn}. The $\UOSP(1|2)$ coefficients are defined as:
\be
|j_3(j_1,j_2),\tau_3(\tau_1,\tau_2);k_3,m_3> \;= \tilde{I}^{j_1}_{(k_1m_1)}{}^{j_2}_{(k_2m_2)}{}_{j_3}^{(k_3m_3)}\; \big[\;|j_1,\tau_1;k_1,m_1>\,\otimes\;|j_2,\tau_2;k_2,m_2>\big].
\ee
where the coefficients are:
\be
\tilde{I}^{j_1}_{(k_1m_1)}{}^{j_2}_{(k_2m_2)}{}_{j_3}^{(k_3m_3)} = \tilde{B}^{j_1j_2j_3}_{k_1k_2k_3}\tilde{C}^{k_1}_{m_1}{}^{k_2}_{m_2}{}_{k_3}^{m_3},
\ee
where $\tilde{C}^{k_1}_{m_1}{}^{k_2}_{m_2}{}_{k_3}^{m_3} := (<k_1,m_1|\;\otimes<k_2,m_2|)|k_3,m_3>$  defines $\SU(2)$ Clebsch-Gordan coefficients and $\tilde{B}^{j_1j_2j_3}_{k_1k_2k_3}$ are factors given by:
\be
\begin{array}{cc}
j_1+j_2+j_3\in\mathbb{N} & j_1+j_2+j_3 \in \mathbb{N}+\frac{1}{2}\vspace{0.2cm}\\

\overbrace{\phantom{xxxxxxxxxxxxxxxxxxxxxxxxxxxxxxxxxxxxxxx}} & \overbrace{\phantom{xxxxxxxxxxxxxxxxxxxxxxxxxxxxxxxxxxxxxxx}} \vspace{0.1cm}\\

\begin{array}{lcr}
\tilde{B}^{j_1j_2j_3}_{j_1j_2j_3} & = & \sqrt{\dfrac{j_1+j_2+j_3+1}{2j_3+1}},\vspace{0.5cm}\\

\tilde{B}^{j_1j_2j_3}_{j_1j_2-\frac{1}{2}j_3-\frac{1}{2}} & = & (-1)^{\lambda_1}\sqrt{\dfrac{-j_1+j_2+j_3}{2j_3}},\vspace{0.5cm}\\

\tilde{B}^{j_1j_2j_3}_{j_1-\frac{1}{2}j_2j_3-\frac{1}{2}} & = & \sqrt{\dfrac{j_1-j_2+j_3}{2j_3}},\vspace{0.5cm}\\

\tilde{B}^{j_1j_2j_3}_{j_1-\frac{1}{2}j_2-\frac{1}{2}j_3} & = & (-1)^{\lambda_1+1}\sqrt{\dfrac{j_1+j_2-j_3}{2j_3+1}},
\end{array}
&

\begin{array}{lcr}
\tilde{B}^{j_1j_2j_3}_{j_1-\frac{1}{2}j_2j_3} & = & (-1)^{\lambda_1+1}\sqrt{\dfrac{-j_1+j_2+j_3 + \frac{1}{2}}{2j_3+1}},\vspace{0.5cm}\\

\tilde{B}^{j_1j_2j_3}_{j_1j_2-\frac{1}{2}j_3} & = & \sqrt{\dfrac{j_1-j_2+j_3+ \frac{1}{2}}{2j_3+1}},\vspace{0.5cm}\\

\tilde{B}^{j_1j_2j_3}_{j_1j_2j_3-\frac{1}{2}} & = & (-1)^{\lambda_1+1}\sqrt{\dfrac{j_1+j_2-j_3+\frac{1}{2}}{2j_3}},\vspace{0.5cm}\\

\tilde{B}^{j_1j_2j_3}_{j_1-\frac{1}{2}j_2-\frac{1}{2}j_3-\frac{1}{2}} & = & \sqrt{\dfrac{j_1+j_2+j_3+\frac{1}{2}}{2j_3}},
\end{array}
\end{array}
\ee
This means that:
\be
\begin{split}
\big[<j_1,\tau_1;k_1,m_1&|\;\otimes<j_2,\tau_2; k_2,m_2|\;\big] \;|j_3(j_1,j_2),\tau_3(\tau_1,\tau_2);k_3,m_3> \\
& =(-1)^{(\lambda_1+2(j_1-k_1))(\lambda_2+2(j_2-k_2)) + \varphi_1 + \varphi_2}\, \tilde{I}^{j_1}_{(k_1m_1)}{}^{j_2}_{(k_2m_2)}{}_{j_3}^{(k_3m_3)}.
\end{split}
\ee
These objects do not have simple transformation properties under permutation.  On the other hand, the $\{3j\}_{\UOSP(1|2)}$-symbols do (by definition):
\bes
 I^{j_1}_{(k_1m_1)}{}^{j_2}_{(k_2m_2)}{}^{j_3}_{(k_3m_3)} &:=& B^{j_1j_2j_3}_{k_1k_2k_3} \,C^{k_1}_{m_1}{}^{k_2}_{m_2}{}^{k_3}_{m_3},\quad\quad\quad\textrm{where:}\\
& \\
B^{j_1j_2j_3}_{k_1k_2k_3} &:=& (-1)^{(\lambda_3+1)(2(j_2 - k_2) + 2(j_1+j_2+j_3)) }\sqrt{2k_3+1}\, \tilde{B}^{j_1j_2j_3}_{k_1k_2k_3},\\
  C^{k_1}_{m_1}{}^{k_2}_{m_2}{}^{k_3}_{m_3} &:=& \frac{(-1)^{k_1-k_2 - m_3}}{\sqrt{2k_3+1}} \tilde{C}^{k_1}_{m_1}{}^{k_2}_{m_2}{}^{k_3}_{-m_3},
\ees
and $C^{k_1}_{m_1}{}^{k_2}_{m_2}{}^{k_3}_{m_3}  $ are the $\{3j\}_{\SU(2)}$-symbols.  Under permutation, they satisfy:
\be
I^{j_{\sigma(1)}}_{(k_{\sigma(1)}m_{\sigma(1)})}  {}^{j_{\sigma(2)}}_{(k_{\sigma(2)}m_{\sigma(2)})}    {}^{j_{\sigma(3)}}_{(k_{\sigma(3)}m_{\sigma(3)})} = (|\sigma|)^{\sum_{a}^3(2(j_a-k_a) (2j_a +1)  + k_a)}I^{j_1}_{(k_1m_1)}{}^{j_2}_{(k_2m_2)}{}^{j_3}_{(k_3m_3)} ,
\ee
where $|\sigma| = \pm1$ is the signature of the permutation and is similar to the $\SU(2)$ case. Reversing the magnetic indices, we find:
\be
I^{j_1}_{(k_1-m_1)}{}^{j_2}_{(k_2-m_2)}{}^{j_3}_{(k_3-m_3)}  =  (-1)^{k_1+k_2+k_3}I^{j_1}_{(k_1m_1)}{}^{j_2}_{(k_2m_2)}{}^{j_3}_{(k_3m_3)} .
\ee
Also, we can raise and lower indices using the invariant metric on the representation space $R^{j,\tau}$ provided by:
\be
\big[<j_1,\tau_1;k_1,m_1|\;\otimes<j_2,\tau_2; k_2,m_2|\;\big] \;|0,0;0,0> = (-1)^{2(j_1-k_1)(2j_1+1) + k_1-m_1} \delta^{j_1j_2}\delta_{k_1k_2}\delta_{m_1\, -m_2}
\ee
This has been mentioned already in \eqref{OSP6}.
Additionally, they satisfy a pseudo-orthogonality relation:
\be
\begin{split}
\sum_{\substack{k_1,k_2\\ m_1,m_2}} (-1)^{(\lambda_1  + 2(j_1-k_1))(\lambda_2 + 2(j_2 - k_2)) + \varphi_1 + \varphi_2}\;  &I^{j_1}_{(k_1m_1)}{}^{j_2}_{(k_2m_2)}{}^{j_3}_{(k_3m_3)} I^{j_1}_{(k_1m_1)}{}^{j_2}_{(k_2m_2)}{}^{j'_3}_{(k'_3m'_3)} \phantom{xxxx} \\
&\phantom{}= (-1)^{\varphi_3(\varphi_1,\varphi_2)}\;\delta^{j_3j'_3}\delta_{k_3k'_3}\delta_{m_3m'_3},
\end{split}
\ee
where $\varphi_3(\varphi_1,\varphi_2) := 2(j_3 - k_3)\rho_3 +\tau_3(\tau_1,\tau_2)$. This implies:
\be
\sum_{\substack{k_1,k_2,k_3 \\ m_1,m_2,m_3}} (-1)^{\Theta}\;    I^{j_1}_{(k_1m_1)}{}^{j_2}_{(k_2m_2)}{}^{j_3}_{(k_3m_3)} I^{j_1}_{(k_1m_1)}{}^{j_2}_{(k_2m_2)}{}^{j_3}_{(k_3m_3)} = 1,
\ee
where:
\be
\Theta := (\lambda_1 + 2(j_1-k_1))(\lambda_2 + 2(j_2 - k_2))  +  2(j_3-k_3)+ \varphi_1 + \varphi_2 + \varphi_3(\varphi_1,\varphi_2).
\ee
As expected, $\Theta$ is invariant under permutation.\footnotemark
\footnotetext{
\be
\begin{array}{rcl}
\Theta & \equiv & \displaystyle \sum_{a = 1}^3(\lambda_a + 2(j_a-k_a))(\lambda_{a+1} + 2(j_{a+1} - k_{a+1})) + \lambda_a\lambda_{a+1}+ 2(j_a - k_a)(\lambda_a + 1),\vspace{0.2cm}\\
& \equiv & \displaystyle \sum_{a = 1}^3 4k_ak_{a+1} + \lambda_a\lambda_{a+1}+ 2(j_a - k_a)(2j_a + 1)
\end{array}
\ee}

\section{Super $\{6j\}$-symbols}
\label{sixj}
The supersymmetric 6j-symbol is defined as the matrix relating two ways of coupling three representations:
\be\label{sixj1}
R^{j_1,\tau_1}\otimes R^{j_2,\tau_2}\otimes R^{j_4,\tau_4} =
\left\{
\begin{array}{l}
\displaystyle \bigoplus_{j_5=|j_3 - j_4 |}^{ j_3+j_4} \bigoplus_{j_3 = |j_1 - j_2|}^{ j_1+j_2} R^{j_5(j_3(j_1,j_2),j_4),\tau_5(\tau_3(\tau_1,\tau_2),\tau_4)}\vspace{0.2cm}\\
\displaystyle \bigoplus_{j_5=|j_1 - j_3 |}^{ j_1+j_3} \bigoplus_{j_3 = |j_2 - j_4|}^{ j_2+j_4}R^{j_5(j_1,j_3(j_2,j_4)),\tau_5(\tau_1,\tau_3(\tau_2,\tau_4))}
\end{array}.\right.
\ee
The states are related by:
\be\label{sixj2}
\begin{split}
&|j_5(j_1,j_6(j_2,j_4)),\tau_5(\tau_1,\tau_6(\tau_2,\tau_4));k_5,m_5> \\
&= \sum_{j_3}(-1)^{(I+1)I_{2,4}+ \lambda_1 I_{2,4}+\lambda_4 I_{1,2} +  \lfloor \frac{I}{2}\rfloor }\left[\begin{array}{ccc}
j_1 & j_2 & j_3\vspace{0.1cm}\\
j_4 & j_5 & j_6
\end{array}\right] |j_5(j_3(j_1,j_2),j_4),\tau_5(\tau_3(\tau_1,\tau_2),\tau_4);k_5,m_5>,
\end{split}
\ee
where:
\be
\begin{array}{cc}

I = 2(j_1+j_2+j_4+j_5), &
\hspace{0.7cm}
\begin{array}{rclrcl}
I_{1,2} & = & 2(j_1+j_2+j_3), &\hspace{0.7cm} I_{1,6} & = & 2(j_1+j_6+j_5),\hspace{0.7cm}\vspace{0.4cm}\\

I_{2,4} & = & 2(j_2+j_4+j_6), & I_{3,4} & = & 2(j_3+j_4+j_5),
\end{array}

\end{array}
\ee
and $\lfloor\frac{I}{2}\rfloor$ is the integer part of $\frac{I}{2}$. This means:\be
\begin{array}{rcl}
\label{susy6j}
\left[\begin{array}{ccc}
j_1 & j_2 & j_3\vspace{0.1cm}\\
j_4 & j_5 & j_6
\end{array}\right]

& := &  \displaystyle\sum_{\substack{k_i,m_i\\1\leq i\leq 6}}(-1)^{ \sum_{a=1}^6[(k_a - m_a) + 2(j_a-k_a)(\lambda_a+ 1)]}\vspace{0.2cm}\\

& &\;\times\; I^{j_1j_2j_3}_{(k_1m_1)(k_2m_2)(k_3m_3)}\;I^{j_5j_6j_1}_{(k_5-m_5)(k_6m_6)(k_1-m_1)}
 I^{j_6j_4j_2}_{(k_6-m_6)(k_4m_4)(k_2-m_2)}\;I^{j_4j_5j_3}_{(k_4-m_4)(k_5m_5)(k_3-m_3)},\vspace{0.4cm}\\

 & = & \displaystyle\sum_{\substack{k_i\\1\leq i\leq 6}}(-1)^{\sum_{a=1}^62(j_a-k_a)(\lambda_a+ 1)}
  B^{j_1j_2j_3}_{k_1k_2k_3}\; B^{j_5j_6j_1}_{k_5k_6k_1}\; B^{j_6j_4j_2}_{k_6k_4k_2}\; B^{j_4j_5j_3}_{k_4k_5k_3}\;
\left\{\begin{array}{ccc}
k_1 & k_2 & k_3\vspace{0.1cm}\\
k_4 & k_5 & k_6
\end{array}\right\}
\end{array}
\ee
The $\SU(2)$ $\{6j\}$-symbol is:
\be\label{sixj5}
\left\{\begin{array}{ccc}
k_1 & k_2 & k_3\vspace{0.1cm}\\
k_4 & k_5 & k_6
\end{array}\right\}
:= (-1)^{\left[\sum_{a = 1}^6 (k_a - m_a)\right]}
 C^{k_1k_2k_3}_{m_1 m_2 m_3}\;C^{k_5k_6k_1}_{-m_5 m_6 -m_1}\;
 C^{k_6k_4k_2}_{-m_6 m_4 -m_2}\;C^{k_4k_5k_3}_{-m_4 m_5 -m_3}\\
\ee
One finds that the supersymmetric $\{6j\}$-symbol has the same symmetry properties as its $\SU(2)$ counterpart.

\section{Diagram evaluation}
\label{diagram}
We have seen in the main text that a spin foam diagram generically factorises upon integration of the group variables and that we can evaluate the amplitude by considering simpler diagrams.  First of all, let introduce some basic elements of the diagrammatic calculus, without justification, for general vector spaces $V,\, W$.  We denote their duals by $V^*,\,W^*$, respectively.  A map $f:V\rightarrow W$ and its dual map $f^*:W^*\rightarrow V^*$ are denoted by:
\begin{figure}[H]\centering
\psfrag{v}{$V$}
\psfrag{w}{$W$}
\psfrag{vd}{$V^*$}
\psfrag{f}{$f$}
\psfrag{fs}{$f^*$}
\includegraphics[width = 8cm]{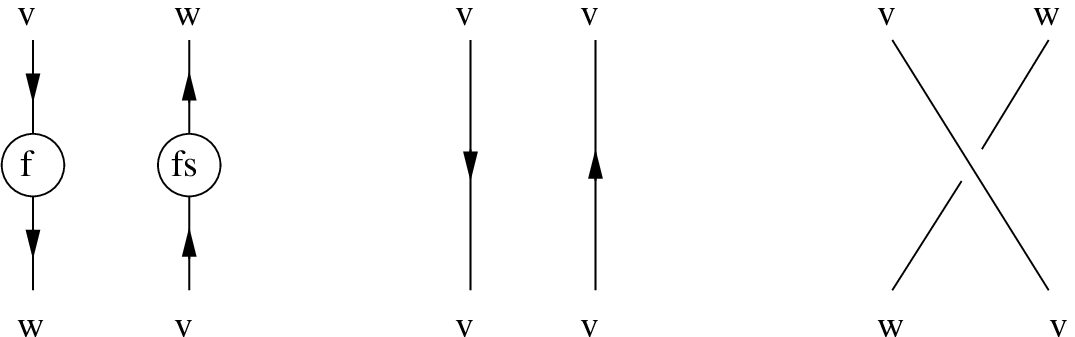}
\end{figure}
\noindent A diagram is always read from top to bottom. We note the direction of the arrows on the diagram, downward arrows map to and from the vector spaces, while upward arrows pass to and from the duals.  Composition of maps follows as one would expect.  The identity maps $\id_V: V\rightarrow V; v\mapsto v$ and $\id_{V^*}:V^*\rightarrow V^*; \phi\mapsto\phi$ are drawn above also.  Finally, we mention the crossing map $\Psi_{V,W}: V\otimes W \rightarrow W\otimes V$.

In our case, we must define evaluation and coevaluation maps along with their duals:
\begin{figure}[H]\centering
\begin{minipage}{0.2\linewidth}\centering
\includegraphics[width=1.5cm]{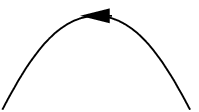}
\end{minipage}
\begin{minipage}{0.5\linewidth}
$coev_V:\mathbb{C}\rightarrow V\otimes V^*\,; \quad1\mapsto \displaystyle\sum_m e_m\otimes f^m,$
\end{minipage}\\

\vspace{0.5cm}

\begin{minipage}{0.2\linewidth}\centering
\includegraphics[width=1.5cm]{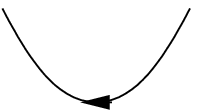}
\end{minipage}
\begin{minipage}{0.5\linewidth}
$ev_V: V^*\otimes V\rightarrow \mathbb{C}\,;\quad  f^m\otimes e_n \mapsto f^m(e_n) = \delta^m{}_n,$
\end{minipage}

\vspace{0.5cm}

\begin{minipage}{0.2\linewidth}\centering
\psfrag{=}{$=$}
\includegraphics[width=3cm]{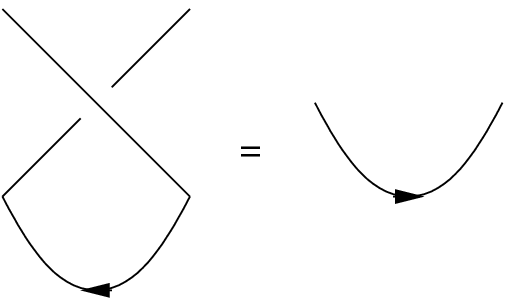}
\end{minipage}
\begin{minipage}{0.5\linewidth}
$coev_V^*: V\otimes V^*\rightarrow \mathbb{C}\,; \quad coev_V^* := ev_v\circ \Psi_{V,V^*},$
\end{minipage}

\vspace{0.5cm}

\begin{minipage}{0.2\linewidth}\centering
\psfrag{=}{$=$}
\includegraphics[width=3cm]{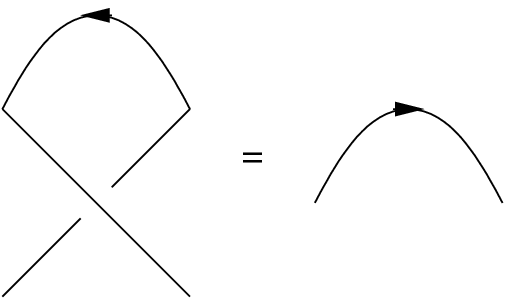}
\end{minipage}
\begin{minipage}{0.5\linewidth}
$ev_V^*: \mathbb{C}\rightarrow V^*\otimes V\,;\quad ev_V^* :=   \Psi_{V^*,V} \circ coev_V,$
\end{minipage}

\end{figure}
\noindent where $e_m$ is a basis for $V$ with $f^m$ its dual basis.

We are interested in othogonal and symplectic vector spaces, that is, vector spaces endowed with either an orthogonal (symmetric, non-degenerate) metric ${}_{o}r$ or a symplectic (anti-symmetic, non-degenerate) metric ${}_{s}r$:

\begin{figure}[H]\centering

\begin{minipage}{0.2\linewidth}\flushright
${}_{o}r: V\otimes V \rightarrow \mathbb{C}\, ; \quad$
\end{minipage}
\begin{minipage}{0.25\linewidth}\centering
\psfrag{r}{${}_{o}r$}
\psfrag{=}{$=$}
\psfrag{-}{}
\includegraphics[width=4cm]{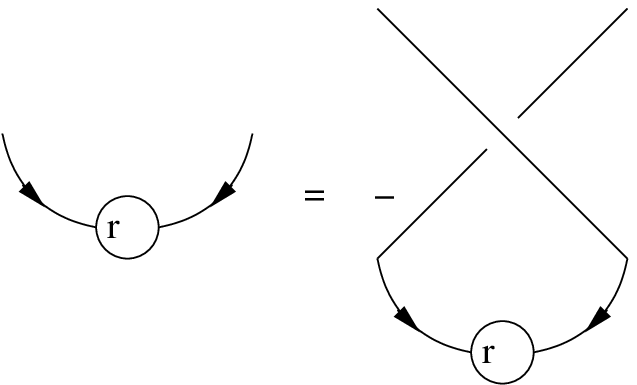}
\end{minipage}
\begin{minipage}{0.2\linewidth}\flushright
${}_{s}r: V\otimes V \rightarrow \mathbb{C}\, ; \quad$
\end{minipage}
\begin{minipage}{0.25\linewidth}\centering
\psfrag{r}{${}_{s}r$}
\psfrag{=}{$=$}
\psfrag{-}{$-$}
\includegraphics[width=4cm]{metric.eps}
\end{minipage}\end{figure}
\noindent where the diagrams demonstrate the (anti-)symmetry.  The existence of a non-degenerate metric allows us to define maps between the vector spaces and their duals, namely, the raising and lowering operators.  The raising operator is given by:
\be
\sharp: V\rightarrow V^*\, ; \quad v \mapsto \left\{ \begin{array}{l} {}_{o}r(v,.),\\ {}_{s}r(v,.),\end{array}\right.
\ee
while the lowering operator $\flat$ is defined such that $\flat \circ \sharp = \id_V$ and $\sharp\circ\flat = \id_{V^*}$.  Thus from a map $f:V\rightarrow W$, we can form another map $f^\flat : V^*\rightarrow W\, ; f^\flat := f\circ\flat$.
Moreover, once such a metric has been defined, there is a natural definition for the crossing map:
\be
\Psi_{V,W}(v \otimes w) = (-1)^{|v||w|} w\otimes v, \quad\quad \textrm{where}\quad\quad |v| , |w| = \left\{ \begin{array}{rl} 0 & \textrm{for $V$ orthogonal,}\\1 & \textrm{for $V$ symplectic.}\end{array}\right.
\ee
Thus, we can now give a more explicit definition of the dual evaluation and coevaluation maps:
\be
{\setlength\extrarowheight{0.2cm}
\begin{array}{rcl}
coev_V^* (e_m\otimes f^n) &=& (-1)^{|e_m||f^n|}\,f^n(e_m) =(-1)^{|e_m||f^n|}\, \delta^n{}_m,\\
ev_V^*(1) &=&\displaystyle \sum_{m} (-1)^{|e_m||f^m|} f^m\otimes e_m
\end{array}}
\ee
Now, let us specialise to the representation spaces of $\SU(2)$ and $\UOSP(1|2)$.  It has been shown in \cite{naish} that in order to obtain a topological state-sum in the $\SU(2)$ case (the Ponzano-Regge model), one must choose the carrier spaces $V^k$, $k\in \mathbb{N}$, to be symplectic and the $V^k$, $k\in\mathbb{N}+\frac{1}{2}$, to be symplectic.  The representation spaces for $\UOSP(1|2)$ fit nicely in with this choice, since the $R^j$ are endowed with an orthosymplectic metric ${}_{os}r$, that is, it is orthogonal on $V^k\subset R^j$, $k\in\mathbb{N}$, and symplectic on $V^k\subset R^j$, $k\in\mathbb{N}+\frac{1}{2}$.    The metric and its for the vector spaces of $\SU(2)$ in their standard bases is:
\be
\begin{array}{rcl}
{}^{k}r^{mn} & = & (-1)^{k - m}\;  \delta^{m+n, 0},\\
{}^{k}r_{mn} & = & (-1)^{ k - n} \; \delta^{m+n, 0}.
\end{array}
\ee
Note that the orthogonal and symplectic nature of the metric is taken care of implicitly in the definition and we can drop the subscripts.  In the course of this work, we must as sum point consider the irreducible representations of $\SU(2)$ within the larger reducible representations $R^j$, and we also may change the inner product on $V^k$ by an overall sign:
\be
\begin{array}{rcl}
{}^{j,k}\tilde r^{mn} & = & (-1)^{2(j - k)(2j+1) + k - m}\;  \delta^{m+n, 0},\\
{}^{j,k}\tilde r_{mn} & = & ((-1)^{2(j - k)(2j+1) + k - n} \; \delta^{m+n, 0}.
\end{array}
\ee
We have already stated the metric for $\UOSP(1|2)$ in \eqref{OSP6}.

\subsection{Simple loop}

\begin{figure}[H]\centering
\begin{minipage}{0.2\linewidth}\centering
\includegraphics[height = 1.5cm]{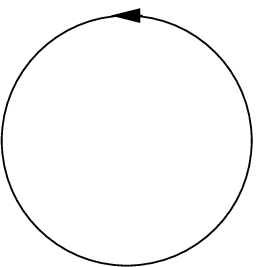}
\end{minipage}
\begin{minipage}{0.2\linewidth}\centering
$= \quad\quad coev_V^* \circ coev_V.$
\end{minipage}
\end{figure}
We are ready to evaluate this diagram in the three different contexts: for $\SU(2)$ with the standard irreducible representations, for $\SU(2)$ with the altered inner product (denoted $\widetilde{\SU}(2)$ in the following); and for $\UOSP(1|2)$.  We shall compute the $\SU(2)$ case explicitly:
\be
coev_{V^k}^* \circ coev_{V^k} :  1\mapsto {}^k e_m \otimes {}^k\! f^m \mapsto \sum_m (-1)^{|{}^ke_m||{}^k\! f^m|}\, {}^k\! f^m({}^ke_m) = (-1)^{2k} (2k+1).
\ee
Completing an identical calculation yields again $(-1)^{2k} (2k+1)$ for $\widetilde\SU(2)$ and $(-1)^{2j}$ for $\UOSP(1|2)$.

Once we start coupled model, we start seeing the appearance of loops on $\widetilde\SU(2)$ with gravity projectors:  ${}^ke_{\pm k}\, {}^k\! f^{\pm k}: v \mapsto {}^ke_{\pm k}\, {}^k\! f^{\pm k}(v) = {}^ke_{\pm k}\, v^{\pm k}$.
\begin{figure}[H]\centering
\begin{minipage}{0.2\linewidth}\centering
\psfrag{+}{$\pm$}
\includegraphics[height = 1.5cm]{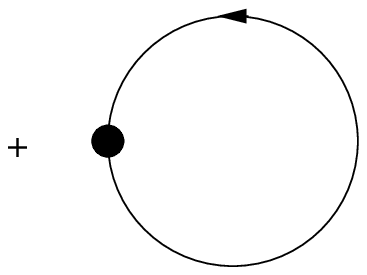}
\end{minipage}
\begin{minipage}{0.5\linewidth}
$= \quad\quad coev_V^* \circ ({}^ke_{\pm k}\, {}^k\! f^{\pm k}\otimes \id_{V^*}) \circ coev_V\,; \quad 1\mapsto (-1)^{2k} $
\end{minipage}
\end{figure}
Moreover, it is a projector so it does not matter how many times it occurs in a loop, the result is the same.

\subsection{Theta}
This amplitude labels the triangles.  This involves the definition of a new map:
\begin{figure}[H]\centering
\begin{minipage}{0.2\linewidth}\centering
\includegraphics[width = 1.5cm]{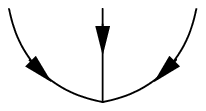}
\end{minipage}
\begin{minipage}{0.5\linewidth}
$C^{k_1k_2k_3}: V^{k_1}\otimes V^{k_2}\otimes V^{k_3}\rightarrow \mathbb{C},$
\end{minipage}
\end{figure}
\noindent with components $C^{k_1}_{m_1}{}^{k_2}_{m_2}{}^{k_3}_{m_3}$. Dualising and applying the lowering operator, one arrives at a map:
\begin{figure}[H]\centering
\begin{minipage}{0.2\linewidth}\centering
\includegraphics[width = 1.5cm]{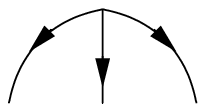}
\end{minipage}
\begin{minipage}{0.5\linewidth}
$\widetilde C_{j_3j_2k_1}: \mathbb{C}\rightarrow V^{k_1}\otimes V^{k_2}\otimes V^{k_3},$
\end{minipage}
\end{figure}
\noindent which, in our cases, we know has components $\widetilde C_{k_3}^{m_3}{}_{k_2}^{m_2}{}_{k_1}^{m_1} = C_{k_3}^{m_3}{}_{k_2}^{m_2}{}_{k_1}^{m_1}$. Thus, the amplitude for the theta diagram is:
\begin{figure}[H]\centering
\begin{minipage}{0.2\linewidth}
\includegraphics[height = 1.5cm]{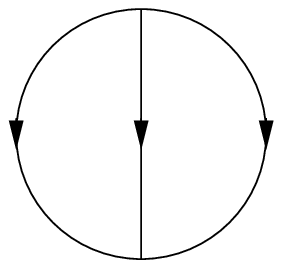}
\end{minipage}
\begin{minipage}{0.7\linewidth}
\be
{\setlength\extrarowheight{0.4cm}
\begin{array}{rrcl}
\SU(2): 			&  C^{k_1}_{m_1}{}^{k_2}_{m_2}{}^{k_3}_{m_3}\,					 C_{k_3}^{m_3}{}_{k_2}^{m_2}{}_{k_1}^{m_1} &=& (-1)^{k_1+k_2+k_3},\\
\widetilde\SU(2):	& 	C^{k_1}_{m_1}{}^{k_2}_{m_2}{}^{k_3}_{m_3}\,					 C_{k_3}^{m_3}{}_{k_2}^{m_2}{}_{k_1}^{m_1} &=& (-1)^{\sum_{a = 1}^3[2(j_a-k_a)+ k_a]},\\
\UOSP(1|2):		&	 I^{j_1}_{(k_1m_1)}{}^{j_2}_{(k_2m_2)}{}^{j_3}_{(k_3m_3)}\,
					I_{j_3}^{(k_3m_3)}{}_{j_2}^{(k_2m_2)}{}_{j_1}^{(k_1m_1)}&=& (-1)^{\lfloor j_1+j_2+j_3\rfloor}.
\end{array}}
\ee
\end{minipage}
\end{figure}
When one embeds the fermion diagrams, some edge amplitudes are no longer labelled by a loop but by a certain type of theta diagram. The following is such an example:
\begin{figure}[H]\centering
\begin{minipage}{0.2\linewidth}
\psfrag{+}{$+$}
\psfrag{-}{$-$}
\includegraphics[height = 1.5cm]{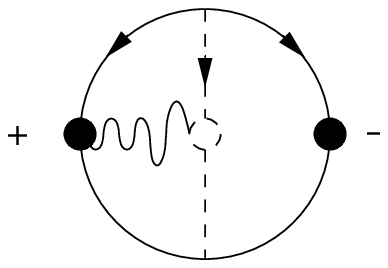}
\end{minipage}
\begin{minipage}{0.7\linewidth}
 \be\nonumber=\quad\quad  (-1)^{2j+1}(2j + 1) \quad
 	C{\begin{array}{rrr} j&\frac{1}{2} &j-\frac{1}{2}\vspace{0.1cm}\\ j&-\frac{1}{2}& -j+\frac{1}{2}\end{array}} \quad
 	C{\begin{array}{rrr} -j+\frac{1}{2}&-\frac{1}{2} &j \vspace{0.1cm}\\ j - \frac{1}{2}&\frac{1}{2}& j \end{array}}
	=  (-1)^{2j}
 \ee
\end{minipage}
\end{figure}
since:
\be
C{\begin{array}{rrr} j&\frac{1}{2} &j-\frac{1}{2}\vspace{0.1cm}\\ j&-\frac{1}{2}& -j+\frac{1}{2}\end{array}}  = (-1)^{2j+1} \sqrt{\frac{1}{2j+1}}.
\ee
So we see that the clasp is just to counteract the factor of $\frac{1}{2k+1}$ in the denominator of the $\{3j\}$-symbol.  There is also further types of diagram contributing to the bosonic sector:\footnote{Remember  that the fermion lines are missing due to the retracing identity.}
\begin{figure}[H]\centering
\begin{minipage}{0.2\linewidth}
\psfrag{+}{$+$}
\includegraphics[height = 1.5cm]{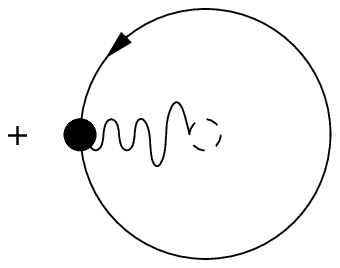}
\end{minipage}
\begin{minipage}{0.2\linewidth}
 \be\nonumber
 =\quad\quad   (-1)^{2(j-k)(2j+1)}(-1)^{2k}(2k+1)
 \ee
\end{minipage}
\end{figure}

\subsection{Tetrahedron}

The tetrahedral diagram turns out to be rather simple contraction of four intertwiners:
\begin{figure}[H]\centering
\begin{minipage}{0.2\linewidth}\centering
\includegraphics[height = 2cm]{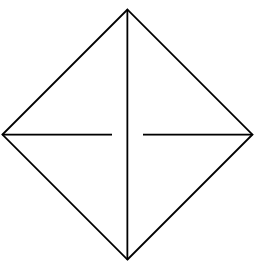}
\end{minipage}
\begin{minipage}{0.7\linewidth}
\be\nonumber
{\setlength\extrarowheight{0.1cm}
\begin{array}{rrcl}
\SU(2): 			&  \left\{\begin{array}{ccc}
k_1 & k_2 & k_3\vspace{0.1cm}\\
k_4 & k_5 & k_6
\end{array}\right\}\\
\widetilde\SU(2):	& 	(-1)^{\sum_{a =1}^6 (2(j_a - k_a )(2j_a+1))}\left\{\begin{array}{ccc}
k_1 & k_2 & k_3\vspace{0.1cm}\\
k_4 & k_5 & k_6
\end{array}\right\}\\
\UOSP(1|2):		&	\left[\begin{array}{ccc}
j_1 & j_2 & j_3\vspace{0.1cm}\\
j_4 & j_5 & j_6
\end{array}\right]
\end{array}}
\ee
\end{minipage}
\end{figure}
Following the trend set so far, when we couple fermionic Feynman diagrams, we find that some triangle amplitudes are in the form of a tetrahedral graph. For example:
\begin{figure}[H]\centering
\begin{minipage}{0.2\linewidth}\centering
\psfrag{a}{${}^{k_2}$}
\psfrag{b}{${}^{k_3}$}
\psfrag{c}{${}^{k_1}$}
\psfrag{d}{${}^{k_2 - \frac{1}{2}}$}
\psfrag{e}{${}^{k_3 - \frac{1}{2}}$}
\includegraphics[height = 2cm]{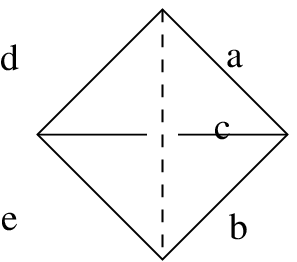}
\end{minipage}
\begin{minipage}{0.7\linewidth}
\be\nonumber = \quad\quad
(-1)^{2k_1 + 2(j_1 - k_1)(2k_1+1)}
\left\{\begin{array}{ccc}
k_1 & k_2 & k_3\vspace{0.1cm}\\
\frac{1}{2} & k_3 - \frac{1}{2} & k_2 - \frac{1}{2}
\end{array}\right\}
\ee
\end{minipage}
\end{figure}
\noindent occurs when we pass from representations in the upper module on the right to the lower on the left.  Furthermore, given that:
\be
	\left\{
	\begin{array}{ccc}
		k_1 & k_2 & k_3\vspace{0.1cm}\\
		\frac{1}{2} & k_3-\frac{1}{2}  & k_2-\frac{1}{2}
	\end{array}
	\right\}
			=	\displaystyle(-1)^{k_1+k_2+k_3}
				\sqrt{\frac{(k_1+k_2+k_3+1)(-k_1+k_2+k_3)}{(2k_3+1)(2k_3)(2k_2+1)(2k_2)}}
\ee
we can see form the numerator that we are starting to get the amplitude $A^{\{j\}}_f(\ua,\ua,\ua; \ua, \da, \da)$ state explicitly in \eqref{susy7}.  There are factors of the square root of various dimensions, but there are factors of dimension multiplying each wedge at the beginning (before integration) and there is exactly the correct factor left over to deal with this denominator.

\section{The integration of three representation functions}
\label{theta}
We mentioned in the main text that there was a non-trivial step in passing from a product of representation functions to intertwiners on the space of representations.  One examines the representation functions occurring in the integral:
\be
\label{tamp1}
A_{e_t^*} =\int_{\UOSP(1|2)} dg_{e_t^*}  \;\;  T^{j_1,\tau_1}_{(k_1m_1),(l_1n_1)}(g_{e_t^*} )\;T^{j_2,\tau_2}_{(k_2m_2),(l_2n_2)}(g_{e_t^*} )\;T^{j_3,\tau_3}_{(k_3m_3),(l_3n_3)}(g_{e_t^*} ),
\ee
for each choice of $k_i$ and $l_i$  as given in (\ref{rep}).  Integrating with respect to $\eta$, $\eta^\square$ and $\Omega$, one should arrive at:
\be
\label{tamp2}
A_{e_t^*} =  I^{j_1}_{(k_1m_1)}{}^{j_{2}}_{(k_2m_2)}{}^{j_{3}}_{(k_3m_3)}I^{j_1}_{(l_1n_1)}{}^{j_2}_{(l_2n_2)}{}^{j_3}_{(l_3n_3)},
\ee
 as stated in the text.  The subtlety becomes clearer when one realises that on the right hand side of  \eqref{rep}, there is no concept of change of $\SU(2)$ module.  Fortunately, there exist relations between the $\SU(2)$ $\{3j\}$-symbols, which provide the missing link between \eqref{tamp2} from \eqref{tamp1}:
\be
\label{tamp4}
\begin{array}{ll}

\left[\left(-j_1+j_2+j_3+\frac{1}{2}\right)\left(j_1-j_2+j_3 +\frac{1}{2}\right)\right]^\frac{1}{2}  C^{j_1-\frac{1}{2}j_2j_3}_{n_1n_2n_3} \vspace{0.4cm} \\

\phantom{xxx}= -\left[\left(j_1+n_1+\frac{1}{2}\right)\left(j_2+n_2\right)\right]^\frac{1}{2}  C^{j_1j_2-\frac{1}{2}j_3}_{n_1+\frac{1}{2}n_2-\frac{1}{2}n_3}

- \left[\left(j_1-n_1+\frac{1}{2}\right)\left(j_2-n_2\right)\right]^\frac{1}{2}C^{j_1j_2-\frac{1}{2}j_3}_{n_1-\frac{1}{2}n_2+\frac{1}{2}n_3},\vspace{0.4cm}\\

\left[\left(j_1+j_2-j_3\right)\left(j_1+j_2+j_3 + 1\right)\right]^\frac{1}{2}  C^{j_1 - \frac{1}{2}j_2 - \frac{1}{2}j_3}_{n_1n_2n_3} \vspace{0.4cm}\\

\phantom{xxx}= \left[\left(j_1+n_1 + \frac{1}{2}\right)\left(j_2-n_2+\frac{1}{2}\right)\right]^\frac{1}{2}  C^{j_1j_2j_3}_{n_1+\frac{1}{2}n_2-\frac{1}{2}n_3}

 - \left[\left(j_1-n_1+\frac{1}{2}\right)\left(j_2+n_2+\frac{1}{2}\right)\right]^\frac{1}{2}C^{j_1j_2j_3}_{n_1-\frac{1}{2}n_2+\frac{1}{2}n_3},\vspace{0.4cm}\\

\left[\left(j_1+j_2-j_3\right)\left(j_1+j_2+j_3 +1\right)\right]^\frac{1}{2}  C^{j_1j_2j_3}_{n_1n_2n_3}\vspace{0.4cm}\\

\phantom{xxx}=  - \left[\left(j_1-n_1\right)\left(j_2+n_2 \right)\right]^\frac{1}{2}  C^{j_1-\frac{1}{2}j_2-\frac{1}{2}j_3}_{n_1+\frac{1}{2}n_2-\frac{1}{2}n_3}

+ \left[\left(j_1+n_1\right)\left(j_2-n_2\right)\right]^\frac{1}{2}C^{j_1-\frac{1}{2}j_2-\frac{1}{2}j_3}_{n_1-\frac{1}{2}n_2+\frac{1}{2}n_3}.
\end{array}
\ee

\end{appendix}


\begin{thebibliography}{99}



  
 \bibitem{naish}
  J.~W.~Barrett and I.~Naish-Guzman,
  {\it The Ponzano-Regge model,}
  Class.\ Quant.\ Grav.\  {\bf 26} (2009) 155014
  [arXiv:0803.3319 [gr-qc]].
  
\bibitem{PR1}
L.~Freidel and D.~Louapre,
  {\it Ponzano-Regge model revisited I: Gauge fixing, observable and interacting spinning particles,}
  Class.\ Quant.\ Grav.\  {\bf 21}, 5685 (2004)
  [arXiv:hep-th/0401076].


\bibitem{livineoeckl}
E.~R.~Livine and R.~Oeckl,
  {\it Three-dimensional quantum supergravity and supersymmetric spin foam models,}
  Adv.\ Theor.\ Math.\ Phys.\  {\bf 7}, 951 (2004)
  [arXiv:hep-th/0307251].


\bibitem{winston1}
  W.~J.~Fairbairn,
  {\it Fermions in three-dimensional spinfoam quantum gravity,}
  Gen.\ Rel.\ Grav.\  {\bf 39}, 427 (2007),
  [arXiv:gr-qc/0609040].

\bibitem{winston2}
R.J. Dowdall and W.J. Fairbairn,
{\it Observables in 3d spinfoam quantum gravity with fermions},
arXiv:1003.1847


\bibitem{YM}
S. Speziale,
{\it Coupling gauge theory to spinfoam 3d quantum gravity},
Class.Quant.Grav.24 (2007) 5139-5160 [arXiv:0706.1534];\\
D. Oriti and H. Pfeiffer,
{\it  A spin foam model for pure gauge theory coupled to quantum gravity},
Phys.Rev.D66 (2002) 124010 [arXiv:gr-qc/0207041].

\bibitem{Livine:2007dx}
  E.~R.~Livine and J.~P.~Ryan,
  {\it N=2 supersymmetric spin foams in three dimensions,}
  Class.\ Quant.\ Grav.\  {\bf 25}, 175014 (2008),
  [arXiv:0710.3540 [gr-qc]].


\bibitem{noncom}
  L.~Freidel and E.~R.~Livine,
  {\it Effective 3d quantum gravity and non-commutative quantum field theory,}
  Phys.\ Rev.\ Lett.\  {\bf 96}, 221301 (2006)
  [arXiv:hep-th/0512113].


  L.~Freidel and E.~R.~Livine,
  {Ponzano-Regge model revisited. III: Feynman diagrams and effective  field theory,}
  Class.\ Quant.\ Grav.\  {\bf 23}, 2021 (2006)
  [arXiv:hep-th/0502106].


\bibitem{gftmat}
W.~J.~Fairbairn and E.~R.~Livine,
  {\it 3d spinfoam quantum gravity: Matter as a phase of the group field theory,}
  Class.\ Quant.\ Grav.\  {\bf 24} (2007) 5277
  [arXiv:gr-qc/0702125].


\bibitem{Scheunert:1976wj}
  M.~Scheunert, W.~Nahm and V.~Rittenberg,
  {\it Irreducible representations of the Osp(2,1) And Spl(2,1) graded Lie algebras,}
  J.\ Math.\ Phys.\  {\bf 18}, 155 (1977).

\bibitem{bertol}
F. A. Berezin and V. N. Tolstoy,
	{\it The group with Grassmann structure UOsp(1.2),}
	Commun. Math. Phys.  {\bf 78}, 409-428 (1981).

\bibitem{Daumens:1992kn}
  M.~Daumens, P.~Minnaert, M.~Mozrzymas and S.~Toshev,
  {\it The superrotation Racah-Wigner calculus revisited,}
  J. Math. Phys. {\bf30} (6), (1993).





\bibitem{aapt}
A.~Achucarro and P.~K.~Townsend,
  {\it A Chern-Simons action for three-dimensional anti-De Sitter supergravity theories,}
  Phys.\ Lett.\  B {\bf 180}, 89 (1986).
\\
A.~Achucarro and P.~K.~Townsend,
  {\it Extended supergravities in d = (2+1) as Chern-Simons theories,}
  Phys.\ Lett.\  B {\bf 229}, 383 (1989).


\bibitem{Bobs}
E.R. Livine and J.P. Ryan,
{\it A Note on B-observables in Ponzano-Regge 3d Quantum Gravity},
Class.Quant.Grav.26 (2009) 035013 [arXiv:0808.0025]


\bibitem{Weinberg}
  S.~Weinberg,
  {\it The Quantum theory of fields. Vol. 1: Foundations,}
  Cambridge, UK: Univ. Pr. (1995).

\bibitem{inprep}
V. Baccetti, M. Dupuis, E.R. Livine and J.P. Ryan,
{\it Asymptotics of the supersymmetric $\{6j\}$-symbol},
in preparation





\bibitem{dewitt}
  B.~S.~DeWitt,
  {\it Supermanifolds,}
 Cambridge, UK: Univ. Pr. (1992) (2nd ed.).













\end{thebibliography}
\end{document}